\documentclass[12pt]{article}

\usepackage{amsmath,amsthm,amssymb,euscript,epsf,epsfig}
\usepackage{mathtools}
\usepackage{array}
\usepackage{fancybox}
\usepackage{comment} 

\usepackage{rotating}

\usepackage{tikz,pbox}
\usetikzlibrary{shapes,arrows}
\usetikzlibrary{calc,decorations.markings}

\usepackage{hyperref}

\makeatletter
\@addtoreset{equation}{section}
\makeatother
\renewcommand{\theequation}{\thesection.\arabic{equation}}

\def\one{{\hbox{ 1\kern-.8mm l}}}
\def\zero{{\hbox{ 0\kern-1.5mm 0}}}

\def\s{ \sigma}

 \def\cK{{\cal K}} 
  \def\cO{{\cal O}}
\def\cP{{\cal P}}  \def\cR{{\cal R}}

\def\Sym{ \hbox{Sym} } 
\def\Orb{ {\rm Orb}} 
\def\Aut{ {\rm Aut} }

\newtheorem{lemma}{Lemma}

\newtheorem{theorem}{Theorem}

\newcommand{\be}{\begin{equation}}
\newcommand{\ee}{\end{equation}}
\newcommand{\beq}{\begin{equation}}
\newcommand{\eeq}{\end{equation}}
\newcommand{\bea}{\begin{eqnarray}\displaystyle}
\newcommand{\eea}{\end{eqnarray}}

\def\s{ \sigma }

\def\Sym{ {\rm{Sym}}  } 
\def\cP{ {\mathcal{P}}}

\begin{document}

\begin{flushright}
QMUL-PH-21-20
\end{flushright}

\bigskip

\begin{center}

{\Large \bf 
All-orders asymptotics of tensor model observables 

from symmetries of restricted  partitions

 }
 \medskip

\bigskip

Joseph Ben Geloun$^{a,c,*}$
 and Sanjaye Ramgoolam$^{b , d ,\dag}  $

\bigskip

$^a${\em Laboratoire d'Informatique de Paris Nord UMR CNRS 7030} \\
{\em Universit\'e Paris 13, 99, avenue J.-B. Clement,
93430 Villetaneuse, France} \\
\medskip
$^{b}${\em School of Physics and Astronomy} , {\em  Centre for Research in String Theory}\\
{\em Queen Mary University of London, London E1 4NS, United Kingdom }\\
\medskip
$^{c}${\em International Chair in Mathematical Physics
and Applications}\\
{\em ICMPA--UNESCO Chair, 072 B.P. 50  Cotonou, Benin} \\
\medskip
$^{d}${\em  School of Physics and Mandelstam Institute for Theoretical Physics,} \\   
{\em University of Witwatersrand, Wits, 2050, South Africa} \\
\medskip
E-mails:  $^{*}$bengeloun@lipn.univ-paris13.fr,
\quad $^{\dag}$s.ramgoolam@qmul.ac.uk

\begin{abstract}
The counting of the dimension of the space of $U(N) \times U(N) \times U(N)$ polynomial invariants of a complex $3$-index tensor as a function of degree $n$ is known in terms of a sum of squares of  Kronecker coefficients. For $n \le N$, the formula can be expressed in terms of a sum of symmetry factors of partitions of $n$ denoted $Z_3(n)$. We derive the large $n$ all-orders asymptotic formula for $ Z_3(n)$ making contact with high order results previously obtained numerically. 
The derivation relies on the dominance  in the sum,  of partitions with many parts of length $1$. The dominance of other small parts  in restricted partition sums leads to related asymptotic results.   The result for the $3$-index tensor observables   gives the large $n$ 
 asymptotic expansion  for the counting of bipartite ribbon graphs with $n$ edges, and for the  dimension of the associated 
 Kronecker permutation centralizer algebra. We explain how the different terms in the asymptotics are associated with  probability distributions over ribbon graphs.  The large $n$ dominance  of small parts  also 
 leads to conjectured formulae for the asymptotics of  invariants for general $d$-index  tensors.  The  coefficients of $ 1/n$ in 
 these expansions involve  Stirling numbers of the second kind along with restricted partition sums.

\end{abstract}

\end{center}

\noindent  Key words: Tensor models, invariant theory,  
asymptotic combinatorics, Kronecker permutation centralizer algebras

\newpage

\tableofcontents

\section{Introduction}

Tensor models are generalizations of random matrix theories where the random variables 
are multi-index tensors. New results on the large $N$ expansion of these models have attracted 
continuing active interest in theoretical physics \cite{Guraubook,Tanasabook}, in particular in connection with 
random discrete geometries \cite{Gurau1102,Bonzom:2011zz}, 
quantum gravity \cite{Rivasseau:2016wvy}, condensed matter  physics, 2D topological field theories \cite{JoSan1}  and models of black hole physics  \cite{Witten:2016iux,KlebanovTasi,Delporte:2018iyf}.  

We will consider  complex tensor variables  $ \Phi_{ i_1 , \cdots  , i_d } $ transforming 
as $ V_N^{ \otimes d } $ under a product of unitary groups $ U(N)^{ \times d } $, 
 where $ V_N$ is the fundamental representation of $U(N)$.  A basis of invariants of $ U(N) ^{ \times d }$ is  built 
using index contractions between $n$ copies of  $ \Phi $ and $n$ copies of  its complex conjugate, 
$ \bar \Phi_{ i_1 , \cdots  , i_d } $ which transforms $\bar V_N^{ \otimes n } $, where $ \bar V_N $ is the complex conjugate representation of $U(N)$. 
The invariant observables are useful as interaction terms in tensor model actions. 
Their enumeration is also of interest in the thermodynamics of quantum mechanical tensor theories
\cite{Tseytlin} and the investigation of their holographic duals \cite{KT1611,Milekhin2008,KMPT1802}.   The leading order large $n$  asymptotics has been discussed in the physics literature in  \cite{Tseytlin,KMPT1802,IMM1710}.

In \cite{JoSan1} the counting and correlators  of the rank-$d$ complex tensor invariants
have been given using permutation equivalences.  A bijection of the permutation basis of invariants 
 with branched covers of the 2-sphere was given and formulations of the counting and correlators 
 in terms of 2D topological field theory were described. 
In \cite{PCAMultMat,JoSan2} it was shown that the invariants at degree $n$ form  a basis 
for an associative algebra $ \cK (n  ) $, denoted the Kronecker permutation centralizer algebra, 
which has a decomposition into blocks of size equal to the Kronecker coefficient for triples of 
Young diagrams with $n$ boxes. The algebra has implications for the structure of tensor model correlators 
\cite{JoSan1,JoSan2}. The algebraic perspectives on tensor model correlators have been developed in 
\cite{DR1706,DGT1707,IY1903}. Similar techniques have been applied to orthogonal invariants 
\cite{Avohou:2019qrl}. Moreover,  in  computational complexity theory
 \cite{MulVar, burg, BCI2011, IkMuWa-VanKron, PakPanova}, Kronecker coefficients 
 form a subject of active interest. 
Very recently, building on the results of \cite{JoSan2}, 
a question of Murnaghan \cite{MurnaghanOnReps, Murnaghan} (discussed among a class of positivity problems in representation theory in \cite{StanleyKron})  about the existence of a  combinatorial  interpretation of the Kronecker coefficient has motivated a construction based on bipartite ribbon graphs  \cite{JoSan3}: for every triple of  Young diagrams with $n$ boxes, the Kronecker  coefficient counts  vectors spanning a
specified  sub-lattice of the lattice of bipartite ribbon graphs of $n$ edges. The vectors are constructed as  null vectors of an integer matrix. The importance of  Kronecker coefficients in mathematics gives  additional  motivations for detailed studies of the properties of $ \cK ( n )$. The asymptotics of the counting of tensor model observables gives the asymptotics of the dimension of $ \cK ( n )$.

The counting formulae in \cite{JoSan1,JoSan2} for $3$-index tensor observables,  in the case $ N \ge n $, 
is recalled as 
\bea\label{Z3form0}  
Z_{ 3 } ( n ) = \sum_{ p \vdash  n } \Sym (p) = \sum_{ R_1 , R_2 , R_3 \vdash n  } C ( R_1  , R_2 , R_3 )^2 
\eea
with $p=(p_1, \dots, p_n)$ a partition of $n$ with symmetry factor $\Sym(p)$, 
and $C ( R_1  , R_2 , R_3 )$ the so-called Kronecker coefficient associated with three Young diagrams $R_1, R_2,$ and $R_3$ with $n$ boxes.  For $ n >  N$, the Young diagrams $R_1,R_2,R_3$  are restricted to have no more than $N$ rows. This counting has also been obtained with motivations from quantum entanglement in 
\cite{HWill,HWW}. The finite $N$ cutoff 
is  a feature  related to Schur-Weyl duality which plays an important role in connection with     the stringy exclusion principle \cite{MalStrom} and giant gravitons \cite{MST}  in the AdS/CFT correspondence \cite{Malda,GKP,Witten} (for a review  of the applications of Schur-Weyl duality in this context see \cite{SWreview}). 
The   asymptotics of \eqref{Z3form0}  at large $n$ has been the subject of a very interesting study  by Kotesovec  up to high order 
using direct numerical fitting techniques and the results are available on the OEIS \cite{VK}. 
In this work, we investigate the asymptotic expansion of
the counting of rank-$d$ tensor invariants. We prove the  asymptotic
expansion series  at all orders for $d=3$. 
Theorem \ref{theosymp} and Theorem  \ref{z3asympt}  are our main results.  We are able to 
match   the series \cite{VK} and extend it  to all orders. We find that the  sum over partitions $p$ in \eqref{Z3form0} 
is dominated by partitions in which most of the parts have  length $1$: these are partitions of the form $[1^{n-k}, q]$ 
where $q$ is a partition of $k$, with $k$ fixed as $n $ tends to infinity. 
Using this dominance  of parts of length $1$, we also conjecture the form of the coefficients of the asymptotic series
for rank $d$ invariants.

This  plan of the paper  is as follows. The next section introduces our notation, 
discusses main features of the asymptotic series  and delivers
our main result, namely the asymptotic expansion
 of the counting of rank $d=3$  of tensor invariants. In the course of this proof, an important 
 role is played by a partition of the set $ \cP ( n )$  of all partitions of $n$ into subsets $\cP_m ( n )$, which are partitions of $n$   where the minimal part length is $m$. Corresponding to these subsets we define $ Z_{ 3 ; m } (n) $. 
 We show that  large $n$ asymptotic series for $ Z_{3} (n ) $ is the same as that for $ Z_{ 3 ; 1 } (n)  $. 
 At the conclusion of this section we explain how the different terms in the asymptotic expansion can be associated with different probability distributions over tensor invariants, or equivalently over bi-partite ribbon graphs. 
 In section \ref{sect:expStirl} we present formulae for the asymptotic series of $ Z_{ 3 ; m } (n)  $, with any finite $m$ which is kept fixed as   $ n $ tends to infinity.   We then 
show that the coefficients of the $1/n$ expansion  can be expressed
in terms of the Stirling numbers of the second kind along with sums over symmetry factors of  restricted partitions.   
Section \ref{sect:hr} elaborates some conjectures about
the expansion for arbitrary rank $d$ invariants, based again on the dominance of small parts. 
A conclusion follows in section \ref{ccl} where we discuss future research directions motivated 
by  this work. 
The paper closes with two  appendices: 
Appendix  \ref{app:proofLemsymq} collects the 
proof of the main lemma which establishes  the dominance
of small cycles, while Appendix  \ref{app:sage}
provides the codes that yield the coefficients
of the asymptotic series expansion of tensor invariants 
at $d=3$ at arbitrary order $n$.

\section{ Asymptotic counting of $3$-index tensor  invariants }

In tensor models we encounter a counting problem involving complex tensor variables 
$ \Phi_{ ijk} $ and $ \bar \Phi^{ i jk }$ which transform in $ V_N^{ \otimes 3 } $ and 
$ \bar V_N^{ \otimes 3}$  of $U(N)$, where $V_N$ is the fundamental of  $U(N)$ and $ \bar V_N$ is the anti-fundamental of $U(N)$. 
The counting of  degree $n$ invariant polynomials when $ n \le  N $ is given by 
\bea\label{Z3form}  
Z_{ 3 } ( n ) = \sum_{ p \vdash  n } \Sym (p) = \sum_{ R_1 , R_2 , R_3 \vdash n  } C ( R_1  , R_2 , R_3 )^2 
\eea
$p$ is a partition of $n$. It is specified by non-negative integers $ (p_1  , p_2 , \cdots , p_n)$ 
which are the numbers of parts of length $1, 2, \cdots , n $. The symmetry factor of the partition is 
\bea 
\Sym ( p ) = \prod_{ i }^{ n } i^{ p_i} p_i! 
\eea
$p$ specifies the cycle structure of a permutation in $ S_n$ : $p_1 $ is the number of cycles of length $1$, $p_2$ is the number of cycles of length $2$, etc.  For a permutation $\sigma $ with cycle structure $p$, $ \Sym ( p ) $ is the number of 
 permutations $ \gamma \in S_n $  satisfying $ \gamma \sigma \gamma^{-1} = \sigma $.  Thus this is the order of the centralizer of 
any $\sigma$ with a given cycle structure $p$.
In the second equality of \eqref{Z3form},
$R_1 , R_2 , R_3$ are Young diagrams with $n$ boxes. $C ( R_1 , R_2 , R_3 )$ is the Kronecker coefficient
for the triple of Young diagrams. The counting has been detailed in \cite{JoSan1, JoSan2}, and has been
generalized to arbitrary rank $d$ tensor $\Phi_{i_1 i_2 \dots i_d}$. 

The asymptotics of $ Z_3 ( n )$ at large $n$ has been calculated by evaluating the sum 
for $n$ up to $20000$ and fitting to the form $ n! P ( 1/n)$ where $P$ is a power series in $1/n$  \cite{VK}: 
\bea\label{Z3Vaclav}
 Z_{ 3} (n ) \sim n! ( 1 + 2/n^2 + 5/n^3 + 23/n^4 + 106/n^5 + 537/n^6  + \cdots ) 
\eea
Remarkably, the coefficients in $P $ are all non-negative integers, this has been verified for the first 
132 terms. 

 This remarkable   integrality is suggestive of some underlying 
 simplicity in the asymptotics. Below we propose a way to 
 understand this simplicity. The idea is to identify a subset of the partitions in \eqref{Z3form} which dominate in the large $n$ limit. 
 
 \subsection{ A partition of the set of partitions  of integer $n$ } 

$ Z_3(n)$ is a sum over partitions of $n$, denoted $p \vdash n $. Let us call this set $ \cP ( n ) $. 
For example $ \cP ( 3 )$ is the set 
\bea 
\cP ( 3 ) = \{ [ 1,1,1] , [ 1,2] , [ 3 ] \} 
\eea
Each square bracket contains positive integers adding to $3$.
The different entries within a bracket are the parts of the partition. 
 We also use the exponent notation 
\bea 
\{ [ 1^3 ], [ 1 , 2] , [ 3 ] \} 
\eea
Each partition of $n$ has integers $i$ with multiplicities $p_i$. 

In order to understand the asymptotics, it will be useful to describe  $\cP ( n )$   as a disjoint union of subsets.
We define $ \cP_{ m } ( n )  $ to be the subset of  $ \cP (n) $ consisting of partitions which have the smallest part equal to 
$ m$. In the above case 
\bea
&& \cP_1 ( 3 ) = \{ [ 1^3] , [ 1, 2 ] \} \cr  
&& \cP_{ 2} ( 3) = \emptyset \cr 
&& \cP_{ 3} ( 3 ) = \{ [3] \}  
\eea
These subsets are evidently disjoint : for any $p$ there is a unique integer $m$ which is the minimum part appearing in $p$. Hence 
\bea 
\cP ( 3) = \cP_1 ( 3) \sqcup \cP_{ 2} ( 3 ) \sqcup \cP_{ 3} ( 3 )   =  \cP_1 ( 3)  \sqcup \cP_{ 3} ( 3 ) 
\eea
For $ n=4$ we have 
\bea 
&& \cP ( 4 ) = \{  [ 1^4] , [ 1^2,2] , [1,3] , [ 2^2]  , [4] \} \cr 
&& \cP_1 ( 4) = \{ [1^4 ] , [ 1^2 , 2] , [1,3] \} \cr 
&& \cP_{ 2} ( 4 ) = \{ [2,2] \} \cr 
&& \cP_{ 3} ( 4 ) = \emptyset \cr 
&& \cP_{ 4 } (4 ) = \{ [4] \} 
\eea
and 
\bea 
\cP ( 4 ) = \cP_{ 1} (4) \sqcup \cP_{ 2} ( 4 ) \sqcup \cP_3 ( 4 ) \sqcup \cP_4 (4)  
 = \cP_{ 1} (4) \sqcup \cP_{ 2} ( 4 ) \sqcup \cP_4 (4)  
\eea
Note that above $m > \lfloor n/2\rfloor$, $ \cP_m ( n) = \emptyset$ unless
$m = n$, and then $ \cP_n( n) = \{ [ n ] \}$. 
In general we have 
\bea\label{decompP}  
\cP ( n ) = \cP_{ 1} ( n ) \sqcup \cP_{ 2} ( n ) \sqcup \cdots \sqcup  \cP_{ \lfloor n/2\rfloor} ( n ) \sqcup \cP_n ( n ) 
\eea

For each of the subsets in the decomposition \eqref{decompP},  we can define the corresponding 
sum over $ \Sym (p)$. First observe that we can write 
\bea \label{Z3n}
Z_{ 3} ( n ) = \sum_{ p \vdash n } \Sym  ( p )  = \sum_{ p \in \cP(n) } \Sym ( p ) 
\eea
For the sums of symmetry factors of partitions restricted to the subsets $ \cP_m (n ) $ of $ \cP ( n )$, we define 
\bea \label{Z3mn}
Z_{ 3 ; m } ( n )  = \sum_{ p \in  \cP_m ( n ) } \Sym ( p )  = \sum_{ \substack { 
p \vdash n \\ p_1 = \cdots = p_{m-1} = 0 } ;\;  p_m > 0  } \Sym ( p ) 
\eea
  The conditions 
$p \vdash n$, $p_1 = \cdots=  p_{m-1} = 0 $ and $p_m > 0$ give an equivalent way to 
express the restriction to $ \cP_m ( n )$. In terms of these restricted partition sums, we have  
\bea \label{Z3nSum}
Z_{ 3 } ( n ) = \sum_{ m =1 }^{ n } Z_{ 3 ; m }  ( n ) \, . 
\eea
It is understood that, for $\lfloor n/2 \rfloor  <  m \le ( n-1)$, $ Z_{ 3 ; m }  ( n )=0$ 
since $ \cP_{ m } ( n )$ is empty in this range, as explained above. 

For the subsequent discussion, it is also useful to define 
\bea 
\cP_{ m^+ } ( n ) = \cP_{ m } ( n )  \sqcup \cP_{ m +1 } ( n ) \sqcup \cdots \sqcup \cP_{ n } ( n ) 
\eea
Thus 
\bea 
 \cP_{ 1^+ } ( n ) &= &  \cP_{ 1} ( n )  \sqcup \cP_2 ( n) \sqcup \cP_3 ( n ) \cdots \sqcup \cP ( n ) \cr 
& = &  \cP ( n ) \cr 
 \cP_{ 2^+} (n) & = &  \cP_2 ( n) \sqcup \cP_3 ( n ) \cdots \sqcup  \cP ( n ) \cr  
& = & \cP  ( n ) \setminus \cP_1 (n )  \cr 
 \cP_{ 3^+ } (  n  ) & = &  \cP_3 ( n ) \sqcup \cP_4 ( n ) \sqcup  \cdots \sqcup  \cP ( n ) \cr  
& = &  \cP  ( n ) \setminus   \{  \cP_1 (n )  \sqcup \cP_2 ( n ) \}  \cr 
& \vdots & \cr 
\cP_{ m^+} ( n )  & =  & \cP ( n ) \setminus \{  \cP_1 (n) \sqcup \cP_2 (n) \sqcup \cdots \sqcup \cP_{ m-1} (  n ) \} 
\eea
\bea 
Z_{ 3 ; m^+ } ( n )  = \sum_{ p \in  \cP_{m^+}  ( n ) } \Sym ( p )  = \sum_{ \substack { 
p \vdash n \\ p_1 = \cdots= p_{m-1} = 0 }   } \Sym ( p ) \,. 
\eea

\subsection{The leading asymptotics} 
Consider some of the terms in the sum \eqref{Z3n}. For $p = [ 1^n]$, $p_1 = n $ and all other $p_i =0$, 
 we have $ \Sym ( p ) = n! $. This very simple 
fact has been used to very good effect, to bound Kronecker coefficients in \cite{PakPanova}. 
Consider another  $ p = [ n ] $, i.e. $ p_n =1 $ and all other $ p_i =0$,
then observe that  $ \Sym ( p ) = n$.  The function $ \Sym (p)$ is such that for a multiplicity $p_i$ 
of cycles, we get factorials of the multiplicity, while the length $i$ of the cycle we get a factor $ i$. 
So large multiplicities of cycles give dominant contributions at large $n$. 
This will be a driving principle in our reasoning.

 In the following, we will prove that the set of all $p$ for fixed $n$ can be organised 
in such a way as to identify the dominant subsets at large $n$. 
The main idea is based on the fact that 
$ Z_3 ( n ) $ is dominated by partitions of  the form $ p = [ 1^{ n-k} , q ]$ of $n$, where $q$ is a partition of $k$ with no parts of length $1$.  
To get a finite order in the asymptotic expansion we need to keep  $k$ finite as $ n$ is taken to infinity. 
Noting that $\Sym( [ 1^{ n-k} , q ] )= ( n -k)! \Sym(q)$, 
consider the finite sum 
\bea \label{z31K}
S_{ 3 , 1 ; K} ( n  ) = 
\sum_{ k =0 }^K  ( n -k)! \sum_{ q \in  \cP_{2^+}  ( k )} \Sym ( q )  
= \sum_{ k =0 }^K  ( n -k)! \sum_{ q \vdash k | q_1 = 0 } \Sym ( q ) 
\eea
The subscript $1$ in $ S_{3 ; 1 ; K } $  indicates that we are isolating the parts of length $1$ and treating the remaining parts as a restricted partition $q$ with no parts of length $1$, that is,  $ q_1 =0$.
Let us illustrate in greater detail that sum. Taking $K = 0$, we have 
\bea 
S_{3  ; 1 ; 0} ( n ) =  \Sym ( [ 1^n] ) = n! 
\eea
Note that $K=1$ does not allow any partition $q$ with $ k=1$, since $k=1$ means there are $n-1$ parts of length $1$, which forces the remaining part to also be of length $1$. 
Let us expand \eqref{z31K} taking $K =4$. 
 we obtain 
\bea \label{z31n}
&& S_{ 3 ; 1 ; 4}  ( n  ) = \crcr
&&  \Sym ( [1^n] ) + \Sym ( [1^{n-2} , 2] ) + \Sym ( [ 1^{ n-3} , 3] ) + \Sym ( [ 1^{ n-4} , 4 ] ) 
 + \Sym ( [ 1^{ n-4}  , 2^2 ] ) \cr 
 && = n! + 2 ( n-2)! + 3 ( n -3)! +  ( n-4)!  ( 4 + 8 ) \cr  
 && = n! \left ( 1 + { 2 \over n ( n-1) }  + { 3 \over n ( n-1) ( n-2) }  + { 12 \over ( n ) ( n -1 ) ( n -2 ) ( n-3) } \right )  \cr 
 && = n! \left (  1 + 2/n^2 + 5/n^3 + 23/n^4  + \cdots \right ) 
\eea
The $k=2$ term gives corrections at order  $1/n^2$ and higher, the $k=3$ gives a correction at order  $ 1/n^3$ and higher. The sum $ S_{ 3 ; 1 ; 4 }$ agrees with the numerical asymptotics of  \eqref{Z3Vaclav}  (from \cite{VK}) up to $ 1/n^4$ corrections. 

We are in a position to formulate our first result. Consider the sum \eqref{z31K} re-expressed in the
following form and the series: for $K\le n$, 
\bea 
&&
S_{ 3; 1  ; K } ( n )  = n! \Big(1 + \sum_{ k =2}^{ K } { 1 \over n ( n-1) \cdots ( n -k+1) } \sum_{ q \vdash k : q_1 =0 } \Sym ( q ) \Big)  \crcr
&&
S'_{ 3;1  ; K } ( n )  = S_{ 3,1  ; K } ( n ) /n! 
\eea
Note that $S'_{ 3,1  ; K } ( n ) $ has an expansion in $1/n$ at large $n$. We
regard this finite sum  as a tool to construct a polynomial in $1/n$ of order $K$, and by taking 
$K$ arbitrarily large and finite (while  $n$ is taken to infinity) we obtain an infinite series
in $1/n$. 
Consider $S_{ 3,1 } ( n )$  as  the sequence of functions $S_{ 3,1  ; K } ( n )$, for all $K$,
and introduce $ S'_{3;1}(n)$ as the sequence $ S_{ 3,1  ; K } ( n )/n!$, for all $K$. 

The following statement holds: 
\begin{theorem}\label{theosymp}
$Z'_{ 3 ; 1 } (n) =  Z_{ 3 ; 1 } (n) / n! $ is asymptotic to $ S'_{3;1}(n)$ in the large $n$ limit. 
\end{theorem}
 
This means we need to show that (see chap. 12.6 \cite{asympt}), 
for fixed and arbitrary $K\ge 0$
\bea 
\lim_{ n \rightarrow \infty } n^K ( Z'_{ 3; 1 } ( n ) - S'_{  3 ; 1; K} ( n ) ) = 0  
\eea
where $S'_{  3 ; 1; K} ( n ) =S_{ 3,1  ; K } ( n )  /n! $.

We write 
\bea\label{TheRemainder}  
&& ( n^K)  ( Z'_{ 3 ;  1 } ( n ) - S'_{ 3,1  ; K } ( n ) )  = { n^{ K } \over n! }    \sum_{  k = K+1 }^{ n } \sum_{ q\vdash k     } \Sym ( [ 1^{n-k} , q ] ) 
\cr 
&&  =   n^K \sum_{ k  = K +1 }^{ n } { 1 \over n ( n -1) \cdots ( n - k  + 1 ) } \sum_{ q \vdash k ; q_1 =0  } \Sym ( q )\equiv \cR_{ n  , K } 
 \eea

It is useful to express the theorem informally as
\begin{equation}\label{informalZ31}  
\boxed{ 
\frac{Z_{  3 ; 1 }( n )}{n!} \sim
\Big(1 + \sum_{ k =2}^{ \infty  } { 1 \over n ( n-1) \cdots ( n -k+1) } \sum_{ q \vdash k : q_1 =0 } \Sym ( q ) \Big)
} 
\end{equation} 
where it is understood that $n$ is being taken to infinity and the precise meaning is the statement following the theorem.

To prove Theorem \ref{theosymp} we need to show that the
remainder $\cR_{ n  , K } $ goes to zero as $ n $ goes to infinity. 
This is the purpose of the next section.

\subsection{ Proof of the asymptotics} 

We first  prove an important fact that partitions { with small parts  have
a dominant  $\Sym(p)$.

 \begin{lemma}
 \label{symq}
 $\forall k \ge 0$ an integer
 \bea
 k \text{ even}\,, 
 &&
 \Sym[2^{k/2}] \ge  \Sym[q],   \qquad q \vdash k, \quad  q_1= 0 
 \label{keven}\\
  k \text{ odd  and }  k\ge 11 \,,   &&
\Sym[3, 2^{(k-3)/2}] \ge  \Sym[q], \qquad q \vdash k, \quad q_1= 0
 \label{kodd}
\eea
 \end{lemma}
 \proof See Appendix \ref{app:proofLemsymq}. 
 
 \qed 
 
\begin{lemma}\label{fnk}
Let $K$ and $n$ be two positive integers, such that
$K \ll n$, and $k \in \{ K, \dots, n \} $. Let $P_1(k)$ be the number of partitions of $k$ with no parts 
 of length $1$. The function $f(n,k)$ defined by 
\bea\label{feo} 
&& 
f ( n , k )  = {  ( n -k) ! \over n! }  P_1( k ) 2^{ k/2} ( k/2) !  \;, 
\quad \text{ for  } k \; \text{ even},   \cr\cr
&& 
f ( n , k )  = {  ( n -k) ! \over n! }  P_1( k ) 2^{ (k-3)/2} ( (k-3)/2) ! 
  \;,  
\quad \text{ for  } k \; \text{ odd},   
\eea
is  maximised by  $f( n , K)$. 
\end{lemma}
\proof 
Let us start with a few comments to explain our proof strategy. 
Numerical investigation shows that the function decreases 
as $k$ increases and reaches a minimum near $ k  = n $. 
In the following, we will estimate the position of the minimal 
as $n$ becomes large.  For $k$ above that minimum, 
 the function increases again but remains at $ k =n$ well below the value at $ k = K$. 

We will prove that the slope of $ f( n , k ) $ as a function of $k$  is negative at $k = K$  and that there 
 is just one minimum in the range $ k = K$ to $ k =n$. Further, we  prove (easily) that the value at $ k = K$  exceeds the value at $ k =  n  $.

When $ k $ is close to $ K $, which is order $1$ as $ n \rightarrow \infty $, then we know that $f$ vanishes,   simply because  $f$ behaves like $ { ( n -K )! \over n! } = O(1/n^{K})$. 
 When $ k $ is close to $n$, i.e. $ ( n - k ) $ is order $ 1 $, then we also know that $ f (  n , k ) $ is vanishing at large $n$. To see this consider  $ f ( n , n )$  in the even case of  \eqref{feo}
 \bea 
 f(n,n) = { 1\over n ! } P_1(n) 2^{ n/2} ( n/2)! \sim { e^{ B  \sqrt n } 2^{ n /2} ( n/2)! \over n ! } \xrightarrow{\text{$n\rightarrow \infty $} }  0 
 \eea
 where we used a standard result for the asymptotics of $ P_1(n)$ where $B$ is a constant, which we give shortly. The odd case is similar. 
 
 We will consider  $ ( n - k )  \sim n^{ \alpha} $ for $ 0 < \alpha < 1$, and thus interpolating between these two limits. 
 For any $ 1 >  \alpha > 0 $, $ k$ and $ ( n - k ) $ both  go to infinity as $ n $ goes to infinity. So we can use  the asymptotic form of $ P_1(k)$. 
The asymptotic behaviour  of $ P_1( k )$ is easily derived from the asymptotics of partition numbers because $ P_1 ( k ) = P ( k ) - P ( k -1)$ (see for example \cite{wenwei}): 
\bea 
\label{p1K}
P_1 ( k ) 
=  { A \over k^{ 3/2} } e^{ B \sqrt { k } }  (1 +  O(\frac{1}{ k^{a}} ) ) 
\eea 
where $A =  \pi /(12 \sqrt{2})$,  $B = \pi \sqrt { 2 /3} $ and $a= 1/2$. 

\ 

\noindent{\bf $k$ even -}
We start by $n$ even and, given the above,  approximate $f(n,k)$ by 
\bea 
f(n , k ) = \frac{1}{n!}\, A k^{ -3/2} e^{ B \sqrt { k } }  (1 + {O \Big( \frac{1}{k^a}\Big) })  \Gamma ( n - k +1 ) 2^{ k/2} \Gamma ( {k \over 2 } + 1 ) 
\eea
As $k$ approaches $n$ for large $n$, the factor  $\Gamma ( n - k +1 )$ decreases 
whereas  $\Gamma ( {k \over 2 } + 1 )$ increases. As we will see, this results in a 
minimum of $f(n,k)$. 

Taking the derivative of $f(n,k)$ with respect to $k$
\bea 
&& { \partial f \over \partial k }  = \crcr
&&
\left ( { - 3 \over 2 k } + { B \over 2  \sqrt{ k}  }  - 
\left ( { 1 \over \Gamma ( z ) } {d \Gamma ( z )  \over dz } \right )\Bigg\vert_{ z = n - k +1 } + { 1 \over 2} \log (2) 
+ { 1 \over 2 }    \left ( { 1 \over \Gamma ( z ) } {d \Gamma ( z )  \over dz }\right )\Bigg \vert_{ z = k/2 +1 }  \right ) f   \cr 
&& 
+  {O \Big( \frac{1}{k^{a+1}}\Big) }
 \frac{1}{n!}\, A k^{ -3/2} e^{ B \sqrt { k } } \,  \Gamma ( n - k +1 ) 2^{ k/2} \Gamma ( {k \over 2 } + 1 ) 
\eea 
The logarithmic derivative of $ \Gamma $ is the di-gamma function 
$\Psi ( z ) = (1/\Gamma ( z )) (d \Gamma ( z ) / dz)$  (also denoted $\psi^{(0)}$). 
We can therefore write 
\bea\label{dfdk} 
&&
{ \partial f \over \partial k }  =\crcr
&&  \Bigg[ \left ( { - 3 \over 2 k } + { B \over 2  k^{ 1/2} }  - 
\Psi ( n - k +1 )  + { 1 \over 2} \log (2) 
+ { 1 \over 2 }   \Psi  (  { k\over 2} +1  )  \right ) \Big(1 + O \Big( \frac{1}{k^{a}}\Big) \Big) 
\cr 
&& 
+  {O \Big( \frac{1}{k^{a+1}}\Big) } \Bigg] 
 \frac{1}{n!}\, A k^{ -3/2} e^{ B \sqrt { k } }  \Gamma ( n - k +1 ) 2^{ k/2} \Gamma ( {k \over 2 } + 1 )
\eea
As noted earlier with  $ ( n - k )  = n^{ \alpha }  $ for $ 0 < \alpha < 1$,  $n - k $ and $k$ both  tend to infinity as $n$ goes to infinity. This is what we are interested in, since the special cases of $ \alpha = 0 , 1 $ are understood by direct calculation. We can  use the asymptotic formula for $ \Psi ( z )$, for any large enough 
$z$: 
\bea 
\Psi ( z ) = \log ( z ) - { 1 \over 2z}   + O\Big(\frac{1}{z^2}\Big)  
\eea

Let us introduce $F(n,k) = \frac{1}{n!}\, A k^{ -3/2} e^{ B \sqrt { k } }  \Gamma ( n - k +1 ) 2^{ k/2} \Gamma ( {k \over 2 } + 1 )$. 
Therefore, we approximate 
\bea 
&&
{ \partial f \over \partial k } = \crcr
&&
\Bigg[
\left( { - 3 \over 2 k } + { B \over 2  k^{ 1/2} } 
+ { 1 \over 2} \log (2) 
  -   \left( \log (  n - k +1 )  - {1 \over  2( n - k +1) } + 
   O\Big(\frac{1}{(  n - k +1 )^2}\Big)   \right)  \right. 
   \crcr
   && \left. 
+ { 1 \over 2 }  
\Big(  \log  ( { k\over 2} +1  )  -  {1\over    k +2   } +  O\Big(\frac{1}{( { k\over 2} +1    )^2}\Big)   \Big) 
 \right)  \Big(1 + O \Big( \frac{1}{k^{a}}\Big) \Big) 
+  {O \Big( \frac{1}{k^{a+1}}\Big) }\Bigg]F(n,k)
\cr\cr
&&
= 
\Bigg[
\Bigg( { - 3 \over 2 k } + { B \over 2  k^{ 1/2} } 
+ { 1 \over 2} \log (2) 
  -   \left( \log (  n - k ) + {1 \over 2 (n-k)}   + 
   O\Big( \frac{1}{( n - k )^2} \Big)   \right) 
   \crcr
   && 
+ { 1 \over 2 }  
\left(  -  \log 2 + \log k  + {1 \over k} +O\Big( \frac{1}{ k^2}\Big)   \right) 
 \Bigg)  \Big(1 + O \Big( \frac{1}{k^{a}}\Big) \Big) 
+  {O \Big( \frac{1}{k^{a+1}}\Big) }\Bigg] F(n,k)
\cr\cr
&&
= 
\Bigg[
\Big( 
 { 1 \over 2 }  \log k  -  \log (  n - k ) -  { 1 \over   k } + { B \over 2  k^{ 1/2} } - {1 \over 2 (n-k)}   
   \crcr
   &&  + 
   O\Big( \frac{1}{( n - k )^2} \Big)   
   +  O\Big( \frac{1}{ k^2}\Big)  
 \Big)  \Big(1 + O \Big( \frac{1}{k^{a}}\Big) \Big) 
+  {O \Big( \frac{1}{k^{a+1}}\Big) }\Bigg]  F(n,k) \, .  
\label{dfdk}
\eea
Given  that $ k , ( n - k ) \rightarrow \infty$, the dominant  terms above in the large $n$ limit  are 
$ { 1 \over 2 }  \log k  -  \log (  n - k ) $. Therefore the condition of vanishing derivative gives 
\bea 
 \log ( n - k ) = { 1 \over 2 } \log ( k )  &\Leftrightarrow & \log ( n - k ) = \log ( \sqrt { k }  ) \cr 
   ( n - k ) = \sqrt { k } & \Leftrightarrow&   k + \sqrt k - n = 0  
\eea  
Solving this quadratic equation for $ \sqrt{ k } $, gives 
$\sqrt{ k  } = { -1 \pm 2 \sqrt{ n } \over 2 } $. 
The positive solution is $ \sqrt{ k } = { -1/2} + \sqrt{ n }$
and yields 
\bea \label{ksqrtn}
k = n - \sqrt{ n } + 1/4
\eea
which confirms $( n - k ) \sim  \sqrt n$ and also $ k^{1/2} <  n-k$.

 \

The above treatment shows that for large $n$ and $( n - k )  \sim  n^\alpha$, $\alpha \in ]0,1[$, 
there is a single extremum of $f(n,k)$, when $k\le n$. 

\ 

\noindent{\bf $k$ odd -} This case can be handled in the similar way as 
above since  
\bea 
&&
{ \partial f \over \partial k } 
= 
\Bigg[
\Big( 
 { 1 \over 2 }  \log k  -  \log (  n - k ) -  { 5 \over  2 k } + { B \over 2  k^{ 1/2} } - {1 \over 2 (n-k)}   
   \crcr
   &&  + 
   O\Big( \frac{1}{( n - k )^2} \Big)   
   +  O\Big( \frac{1}{ k^2}\Big)  
 \Big)  \Big(1 + O \Big( \frac{1}{k^{a}}\Big) \Big) 
+  {O \Big( \frac{1}{k^{a+1}}\Big) }\Bigg]  \tilde F(n,k)  
\eea
with 
$\tilde F(n,k) = \frac{1}{n!}\, A k^{ -3/2} e^{ B \sqrt { k } }  \Gamma ( n - k +1 ) 2^{ (k-3)/2} \Gamma ( {(k-3) \over 2 } + 1 )$, which departs 
\eqref{dfdk} by an irrelevant term. Indeed, $-5/2k$
does not contribute in the remaining analysis and we arrive
at  the exact same result in the same approximation.
Thus, \eqref{ksqrtn} holds in both 
regimes $k^{1/2} \le n-k$ and $k^{1/2} > n-k$. 
 
\ 

Appendix \ref{app:sage} gathers numerical evaluations of  the minimum 
value of $f(n,k)$. It shows that the approximation $\sqrt{ k_{\min} } \sim n -k_{ min} $
holds for a range of values of $n$.  We have set $ n \in [20, 80]$ for $n$
even, and $n\in [21, 81]$, for $n$ odd.

\

Now we address the slope of the function at $k=K$ and at $k=n$. 

\noindent{\bf Slope at $K$.} We compare $f(n,K)$ and $f(n,K+2)$,
(notice that $K$ and $K+2$ share the same parity)
at large $n \gg K$, 
\bea
\frac{f(n,K)}{f(n,K+2)} = 
(n-K) G(K)  \ge  1  
\eea
with a finite function $G(K)$. Thus $f$ is decreasing at $K$. 

\ 

\noindent{\bf Slope at $n$.} Consider $n$ even, 
we compare $f(n,n)$ and $f(n,n-2)$, at large $n \gg K$.  
We get 
\bea
\frac{f(n,n)}{f(n,n-2)} = \frac{1}{2}
\frac{P_1(n)}{P_1(n-2)} \frac{2^{n/2}}{2^{(n-2)/2}} 
\frac{(n/2)!}{((n-2)/2)!}  
= \frac{P_1(n)}{P_1(n-2)}(n/2) \ge 1 \,, 
\eea
with $P_1(n) \ge P_{1}(n-1)$. 
Thus $f$ is increasing between $n-2$ and $n$.

On the other hand, if $n$ odd and large, we have 
 \bea
\frac{f(n,n)}{f(n,n-2)} = \frac{1}{2}
\frac{P_1(n)}{P_1(n-2)} \frac{2^{(n-3)/2}}{2^{(n-5)/2}} 
\frac{((n-3)/2)!}{((n-5)/2)!}  
= \frac{P_1(n)}{P_1(n-2)}((n-3)/2) \ge 1 \,, 
\eea
Again, w conclude tha $f$ increases 
between $n-2$ and $n$.

\

\noindent{\bf $f(n,K)$ is the max.}
The last piece of information we need is the comparison 
between $f(n,K)$ and $f(n,n)$. As $K$ is a finite small 
integer, we want to show that $f(n,K) \ge f(n,n)$
at large $n$. We note that $k=K, \dots, n$
could be even or odd, and therefore we 
could only compare $K$ and $n$ having 
the same parity. 

Assuming that $K$ and $n$ are even,
for $n$ large and finite $K$,  
using Stirling approximation and \eqref{p1K},
we have 
\bea
&&
\frac{f(n,K)}{f(n,n)} = (n-K)! 
\frac{P_1(K)}{P_1(n)} \frac{2^{K/2}}{2^{n/2}} 
\frac{(K/2)!}{(n/2)!}  \crcr
&&
 = G(K) 
 \frac{\sqrt{2 \pi (n-K)} \Big(\frac{(n-K)}{e}\Big) ^{n-K}  \Big(1 + O(\frac1n)\Big) }{
  { A \over n^{ 3/2} } e^{ B \sqrt { n } }  (1 +  O(\frac{1}{ n^{1/2}} ) )  2^{n/2} 
   \sqrt{ \pi n} \Big(\frac{n}{2e}\Big) ^{n/2}  \Big(1 + O(\frac1n)\Big) 
  }\crcr
&&
 = G_1 (K) 
 \frac{ \sqrt{2 (n-K)} n ^{n/2-K } (n(K-1) + K^2  + O(\frac{1}{n}))  }{
  A e^{ B \sqrt { n }  + n/2    }  (1 +  O(\frac{1}{ n^{1/2}} ) ) 
  }  \ge 1
\eea
because, at large $n$ and finite $K$, $n^{n/2}$ dominates
$e^{ B \sqrt { n }  + n/2    } $.  Above $G(K)$ and $G_1(K)$ are finite
functions of $K$. 
In the same vein, considering that $K$ and $n$ are odd,
we have 
\bea
\frac{f(n,K)}{f(n,n)} = (n-K)! 
\frac{P_1(K)}{P_1(n)} \frac{2^{(K-3)/2}}{2^{(n-3)/2}} 
\frac{((K-3)/2)!}{((n-3)/2)!}   \ge 1
\eea
and the inequality can be justified once 
by exactly the same argument 
at large $n$ and finite $K$. 
This ends the proof of the lemma.

\qed 

Theorem \ref{theosymp} is a straightforward corollary of on the following
statement.
\begin{theorem}
\label{remainderRk}
The remainder \eqref{TheRemainder} obeys the limit:
$\lim_{n \to \infty}
 \cR_{ n , K } =  0$. 
 \end{theorem}
\proof  Let us separate the remainder into two parts: 
\bea 
\cR_{ n  , K } = \cR_{ n , K }^+ + \cR_{ n , K }^-   
\eea
where $ \cR_{ n , K }^+$ is a sum over even $k$ and $ \cR_{ n , K }^-$ is a sum over odd $k$. 
To write explicit formulae for these, we will need to treat $K$ even and odd separately. 

\

\noindent{\bf $K$ is odd.} We seek  an upper bound for
\bea 
\cR_{ n , K }^+  =  n^K \sum_{ k = K+1 ;  ~   k ~  \text{even}}^{n } 
  {  ( n -k) ! \over n! } \sum_{ \substack{ q \vdash k \\q_1 =0}  } \Sym ( q ) 
\eea
Lemma \ref{symq} shows that $\Sym(q)$ is maximised by $ q = [ 2^{ k/2} ] $
for $k$ even. The number of terms in the sum over $q$ is $P_1(k)$. Hence, we claim 
 \bea 
 \cR_{ n , K }^+  < n^K \sum_{ k = K+1 ;  ~   k ~  \text{even} }^{n }   {  ( n -k) ! \over n! }  P_1( k ) 2^{ k/2} ( k/2) ! \equiv 
 \tilde \cR_{ n , K }^+  
 \eea
It is convenient to separate $ \tilde \cR_{ n , K }^+$ into two parts: the first term in the sum and the rest. 
\bea\label{tildecRsm} 
&& \tilde \cR_{ n , K}^+ = n^K { ( n -K -1 )! \over n! } P_1 ( K +1) 2^{ (K +1 )/2} ( (K+1)/2)! \cr 
&& +  n^K \sum_{ k = K +3 ;  ~   k  ~    \text{even}  }^{ n }  {  ( n -k) ! \over n! }  P_1( k ) 2^{ k/2} ( k/2) ! 
\eea
For large $n$ and $ K $ finite, the first term can be expressed as 
\bea
&&
 n^K { ( n -K -1 )! \over n! } P_1 ( K +1) 2^{ (K +1 )/2} ( (K+1)/2)!  
=   C_K  n^K { ( n -K -1 )! \over n! } \crcr
&&
= C_K n^K \frac{1}{n (n-1) \dots (n-K)}
 = C_K n^K\frac{1}{n^{K+1}}  (1+ O\Big(\frac{1}{n}\Big))
 =O\Big(\frac{1}{n}\Big)
\eea
for $C_K$ a constant. 
The first term scales  as $1/n$, so vanishes at large $n$.

Lemma \ref{fnk} implies that the summand in the second term  of \eqref{tildecRsm} is maximised by the first term of that sum provided that $ n \gg K+3$. Therefore, the sum in \eqref{tildecRsm} is bounded from 
above by 
\bea 
n^K { ( n - K -3 + 2 ) \over 2 }  { ( n - K - 3 ) ! \over  n! } P_1 ( K +3 ) 2^{ ( K +3 ) /2 }   ({ (  K +3 ) / 2 })  !
\eea 
The factor $ ( n - K -3 + 2 )/2  $ bounds from above the number of terms in the sum over $k$ (for $n$ even it is exact, for $n$ odd it exceeds
the number of terms by $1/2$). At large $n$ and fixed  $K +3 \ll n$, 
we expand the previous expression as
\bea 
C_K\,   n^K  ( n - K )  { ( n - K - 3 ) ! \over  n! }  = 
 C'_K\,   {n^{ K+1} \over n^{ K +3} }  (1+ O\Big(\frac{1}{n}\Big)) 
  =  O\Big(\frac{1}{n^2}\Big)
\eea
for some finite positive constants $C_K, C_K'>0$. 
This goes to zero at large $n$. 

\

\noindent{\bf $K$ is even.}   Although in the following we keep the same notation,
the reader should be aware that the expressions may
designate different quantities. 
We adopt the same strategy 
as above and find upper bound for the remainder 
and show that it goes to 0 as $n$ tends to infinity. 
We have
\bea 
\cR_{ n , K }^+  =  n^K \sum_{ k = K+2 ;  ~   k ~  \text{even}}^{n } 
  {  ( n -k) ! \over n! } \sum_{ \substack{ q \vdash k \\q_1 =0}  } \Sym ( q ) 
\eea
Once again,  $k$ is even, so Lemma \ref{symq} provides 
a bound on $\cR_{ n , K }^+$ as  follows 
 \bea 
 \cR_{ n , K }^+  < n^K \sum_{ k = K+2 ;  ~   k ~  \text{even} }^{n }   {  ( n -k) ! \over n! }  P_1( k ) 2^{ k/2} ( k/2) ! \equiv 
 \tilde \cR_{ n , K }^+  
 \eea
 The rest of the proof is similar: we separate the 
 first term $n^K  {  ( n -K-2 ) ! \over n! }  P_1( K+2  ) 2^{ (K+2 )/2} ( (K+2 )/2) ! 
 = O(n^{-2})$ and the remaining partial sum assumes the bound, by Lemma \ref{fnk}, 
\bea
C_K n^K \frac{(n - (K+4) +2 )}{2}    {  ( n -K - 4) ! \over n! }  = O\Big(\frac{1}{n^{3}}\Big)
\eea
for $C_K>0$ a constant. 

\

We now concentrate on the sum over $k$ odd
in the remainder. The above routine allows us to prove the statement. 

\noindent{\bf $K$ is odd.} Consider the remainder 
\bea 
\cR_{ n , K }^-  =  n^K \sum_{ k = K+2 ;  ~   k ~  \text{odd}}^{n } 
  {  ( n -k) ! \over n! } \sum_{ \substack{ q \vdash k \\q_1 =0}  } \Sym ( q ) 
\eea
For this case, Lemma \ref{symq} leads  us to the bound
 \bea 
 \cR_{ n , K }^-  < n^K \sum_{ k = K+2 ;  ~   k ~  \text{odd} }^{n }   {  ( n -k) ! \over n! }  P_1( k ) \,  3 \cdot 2^{ (k-3)/2} ( (k-3)/2) ! \equiv 
 \tilde \cR_{ n , K }^- 
 \eea
 The first term $ n^K   {  ( n -K-2) ! \over n! }  P_1( K+2 ) \,  3 \cdot 2^{ (K-1)/2} ( (K-1)/2) ! $ behaves like $O(1/n^{2})$, and, still by Lemma \ref{fnk},
 we treat the remaining partial  sum  by  the bound 
 \bea
  n^K \frac{(n- K-2)}{2 } {  ( n -K-4) ! \over n! }  P_1( K+4 ) \,  3 \cdot 2^{ (K+1)/2} ( (K+1)/2) !  =  O\Big(\frac{1}{n^{3}}\Big) 
 \eea

\noindent{\bf $K$ is even.} This is the last case to deal with. 
We express the remainder as 
\bea 
\cR_{ n , K }^-  =  n^K \sum_{ k = K+1 ;  ~   k ~  \text{odd}}^{n } 
  {  ( n -k) ! \over n! } \sum_{ \substack{ q \vdash k \\q_1 =0}  } \Sym ( q ) 
\eea
Lemma \ref{symq} gives us 
 \bea 
 \cR_{ n , K }^-  < n^K \sum_{ k = K+1 ;  ~   k ~  \text{odd} }^{n }   {  ( n -k) ! \over n! }  P_1( k ) \,  3 \cdot 2^{ (k-3)/2} ( (k-3)/2) ! \equiv 
 \tilde \cR_{ n , K }^- 
 \eea
 The first term $ n^K   {  ( n -K-1) ! \over n! }  P_1( K+1 ) \,  3 \cdot 2^{ (K-2)/2} ( (K-2)/2) ! $ behaves like $O(1/n)$, and,  Lemma \ref{fnk} bounds the remaining partial  sum with 
 \bea
  n^K \frac{(n- K-1)}{2 } {  ( n -K-3) ! \over n! }  P_1( K+3 ) \,  3 \cdot 2^{ K/2} ( K/2) !  =  O\Big(\frac{1}{n^{2}}\Big) 
 \eea
This ends the proof of the theorem. 

\qed

The following statement holds
\begin{theorem}\label{z3asympt}
$Z_{ 3 } (n) $  is asymptotic to $ S_{ 3 ; 1 } ( n )$ in the large $n$ limit. 
\end{theorem}
 As in \eqref{informalZ31} it is useful to express this result informally as 
\begin{equation}\label{informalZ3}  
\boxed{ 
Z_{  3  } ( n ) \sim n!
\Big(1 + \sum_{ k =2}^{ \infty  } { 1 \over n ( n-1) \cdots ( n -k+1) } \sum_{ q \vdash k : q_1 =0 } \Sym ( q ) \Big)
} 
\end{equation}

\proof[Proof of Theorem \ref{z3asympt}]
We  show that, $Z'_3(n)= Z_3(n)/n!$, for all $K$
\bea
\lim_{n\to \infty} n^K (Z'_3(n) - S'_{3;1;K}(n)) = 0 
\eea
Note first that $Z_3(n)$ expands as
\bea
Z_3(n) = Z_{3;1} (n)  + Z_{3;2^+} (n)
\eea
$Z_{3;2^+} (n)$ sums over partitions with no parts of size 1. 
Lemma \ref{symq}  teaches that, for all $n \ge 0$,  and  even or odd, 
the maximal $\Sym[q]$,  among $q \vdash n$, with $ q_1= 0 $, 
is known.

-  Let us focus on the case $n$ even: 
\bea
Z_{3;2^+} (n) \le \Sym[2^{n/2}] P_1(n)
\eea
where $P_1(n)$ keeps its previous meaning as the number
of the partitions of $n$ without parts of size 1. It becomes obvious that
\bea
Z_3(n) \le Z_{3;1} (n)  +  \Sym[2^{n/2}] P_1(n)
\eea
We rewrite this under the light of Theorem \ref{theosymp}: 
\bea
n^K \Big(Z'_3(n) - S'_{3;1;K}(n) \Big) 
 &\le& n^K \Big(  Z'_{3;1} (n)  - S'_{3;1;K}(n)   + \frac{1}{n!}  \Sym[2^{n/2}] P_1(n)    \Big)
   \crcr
  &\le&  \cR_{n,K}   + \frac{n^K}{n!}  \Sym[2^{n/2}] P_1(n)  
\eea
Taking the limit when $n\to \infty$, 
using Theorem \ref{remainderRk} showing $ \cR_{n,K} \to  0$ and,
the fact that $n^K \Sym[2^{n/2}] P_1(n)   \sim   n^{n/2 +  K-3/2} e^{B \sqrt{n}-n/2}$ 
is suppressed by the denominator $n! \sim n^n e^{-n}$, for any finite $K$, 
we obtain the result. 

-  In the same vein, when $n$ is odd, we have for $n> 11$ \eqref{kodd}, 
\bea
Z_{3;2^+} (n) \le   \Sym[3, 2^{(n-3)/2}]  P_1(n)
\eea
which leads, using the same above argument, to a vanishing 
remainder. 

\qed 

\noindent
{\it Remark:}   Note that $ Z_{ 3 ; 1} ( n ) $ and $ Z_3 (n)$ have the same large $n$ asymptotic expansion. 
We will discuss a large $n$ characterization of $ Z_3( n)$ which conjecturally fixes it uniquely in section \ref{sec:NPA}.

\subsection{ Interpretation in terms of probability distributions  over bipartite ribbon graphs } 

It is natural to ask how we should interpret the asymptotic results we have found here. 
Which tensor invariants dominate in the large $n$ limit? As explained in \cite{JoSan1,JoSan2,JoSan3} 
the tensor invariants of degree $n$ correspond to bi-partite ribbon graphs with $n$ edges. 
These are in 1-1 correspondence with orbits of an action by $ S_n$ on   pairs $ ( \sigma_1 , \sigma_2) \in S_n \times S_n$. The action of $ \gamma \in S_n$ is given by  
\bea 
( \sigma_1  , \sigma_2 ) \sim ( \gamma \sigma_1 \gamma^{-1} , \gamma\sigma_2 \gamma^{-1} ) 
\eea
For each orbit there is a tensor invariant or bi-partite ribbon graph. Letting $r$ be an index running over the set of bipartite ribbon graphs, we can pick pairs  $ ( \sigma_1^{(r)} , \sigma_2^{(r)} ) \in S_n \times S_n  $ in the orbit.  One way to understand asymptotic results is to find configurations that dominate. For example the Plancherel distribution for Young diagrams is dominated by typical Young diagrams with shape close to a limit curve \cite{VershikKerov}. So is there a class of bi-partite graphs which dominate  in the large $n$ limit? The short answer is that rather than dominant ribbon graphs, the explanation that follows from the derivation is that the leading asymptotics is determined by a probability distribution over ribbon graphs. 

To understand this,  recall the derivation using Burnside Lemma  
\bea\label{Z3formderive} 
 Z_3 (n) &= &  { 1 \over n! } \sum_{ \gamma \in S_n } \sum_{ \sigma_1 , \sigma_2\in S_n  } \delta ( \gamma \sigma_1 \gamma^{-1} \sigma_1^{-1} )   \delta ( \gamma \sigma_2 \gamma^{-1} \sigma_2^{-1} )\cr 
&&  = { 1 \over n! } \sum_{ p \vdash n } { n! \over \Sym  (p)  } ( \Sym (p)  )^2  = \sum_{ p } \Sym (p) 
\eea 
which is given and explained in more detail  in \cite{JoSan1}. 
The factor $ { n! \over \Sym (p)  } $ is the number of permutations $ \gamma $ in the conjugacy class $ p$. 
The two sums over $ \sigma_1 , \sigma_2$ give the factor $ ( \Sym (p)  )^2 $. 
Alternatively we can take the sum over $ ( \sigma_1 , \sigma_2 )$ outside and write it as a sum over orbits. We use 
$ |\Orb ~ ( r )  | $ to denote  the number of permutation pairs in the orbit of $ ( \sigma_1^{(r)} , \sigma_2^{(r)} ) $ and $ |\Aut  ~ ( r)  |$ is the number of permutations $ \gamma $ leaving fixed the pair  $ ( \sigma_1^{(r)} , \sigma_2^{(r)} ) $.  By the orbit stabilizer theorem we have $ n!/ |\Orb ~ ( r )  | =  |\Aut  ~ ( r )  |$.
We will use $ \Aut ~ ( r )  \cap [p] $ to denote the subset of $ \gamma \in \Aut ~  ( r )  $ which belong to the conjugacy class $[p]$ where the cycles of $ \gamma$ define the partition $p$ of $n$ 
(we denote this as $ [\gamma] = [p]$ below):
\bea 
Z_3 ( n ) & = &  \sum_{ \sigma_1 , \sigma_2\in S_n }    { 1 \over n! } \sum_{ \gamma \in S_n } \delta ( \gamma \sigma_1 \gamma^{-1} \sigma_1^{-1} )   \delta ( \gamma \sigma_2 \gamma^{-1} \sigma_2^{-1} ) \cr 
 & = & \sum_{ r } { 1 \over n! }  {  | \Orb ~( r )  | } \sum_{ \gamma } 
 \delta ( \gamma \sigma_1^{(r)}  \gamma^{-1} )( \sigma_1^{(r)})^{-1} )  
  \delta ( \gamma \sigma_2^{(r)}  \gamma^{-1} (\sigma_2^{(r)})^{-1} )  \cr 
  & = & \sum_{ r } { 1 \over n! }  {  | \Orb ~ ( r )  | } \sum_{ p  \vdash n } 
  \sum_{ \gamma : [\gamma ] = [p] }  \delta ( \gamma \sigma_1^{(r)}  \gamma^{-1} )( \sigma_1^{(r)})^{-1} )  
  \delta ( \gamma \sigma_2^{(r)}  \gamma^{-1} (\sigma_2^{(r)})^{-1} ) \cr 
  & = & \sum_r { 1 \over |\Aut ~(  r ) |  } \sum_{ p \vdash n } |\Aut ( r ) \cap [p] | \cr 
  & = & \sum_{ p \vdash n }   \sum_r { 1 \over |\Aut ~ (  r ) |  } |\Aut ( r ) \cap [p] | 
\eea
For each fixed $p$, the sum over $ r $ gives $ \Sym (p)$
\begin{equation}\label{symexprib} 
\boxed{ ~~~~~~~~ 
\Sym (p) =  \sum_r { |\Aut ( r ) \cap [p] | \over |\Aut ~ ( r) |  }
~~~~~~~~~} 
\end{equation} 
from which the formula \eqref{Z3formderive} for $ Z_3 ( n )$ as a sum of the symmetry factors. This is a very interesting equation. The LHS is an integer defined entirely in terms of $S_n$. For any $\sigma $ in the conjugacy class $[p]$, it is the number of permutations $ \gamma \in S_n$ such that $ \gamma \sigma \gamma^{-1} = \sigma $. On the RHS we have a sum over bipartite ribbon graphs with $n$ edges (equivalently over tensor invariants). Each term is a positive rational number smaller or equal to $1$.  It is useful to spell out the derivation of \eqref{symexprib}:
\bea 
&& \sum_{ r }  { |\Aut ( r ) \cap [p] | \over |\Aut ~ ( r) |  } 
= \sum_{ r }  { | \Orb ( r ) | \over n! } |\Aut ( r ) \cap [p] | \cr 
&&  = \sum_{ r } {| \Orb ( r ) | \over n! } \sum_{ \gamma : [\gamma ] = [p] }  \delta ( \gamma \sigma_1^{(r)}  \gamma^{-1} )( \sigma_1^{(r)})^{-1} )  
  \delta ( \gamma \sigma_2^{(r)}  \gamma^{-1} (\sigma_2^{(r)})^{-1} ) \cr 
  && = \sum_{ \s_1 , \s_2 \in S_n } \sum_{ \gamma : [\gamma ] = [p] } { 1 \over n! } \delta ( \gamma \sigma_1 \gamma^{-1} \sigma_1^{-1} )   \delta ( \gamma \sigma_2 \gamma^{-1} \sigma_2^{-1} )  \cr 
  && = \sum_{ \gamma : [\gamma ] = [p] } { 1 \over n! } ( \Sym (p ) )^2
   = { n! \over \Sym (p ) } { 1 \over n! } ( \Sym (p ) )^2 \cr 
  && = \Sym ( p ) 
\eea
We used the fact the number of permutations in the class $[p]$ is  ${ n! \over \Sym (p ) }$ and the structure of the proof is essentially reversing, at fixed $p$, the steps of \eqref{Z3formderive}.

The equation \eqref{symexprib} means that, for each $p$,  we can define a probability distribution $ W ( p , r ) $ over ribbon graphs 
\bea 
W ( p , r ) = { 1 \over \Sym ( p )  } { |\Aut ( r ) \cap [p] | \over |\Aut ~ ( r ) |  } 
\eea 
Since our asymptotic results have been derived by organising  the set of $p$ in the sum for $ Z_3 (n)$ according to powers of $n$, each term can be interpreted using $W( p,r)$.  Taking $p = [1^n]$ which contributes the leading term in the asymptotics of $ Z_3 ( n )$, the equation \eqref{symexprib} becomes   
\bea 
n! = \sum_{ r }  { 1 \over |\Aut ~ r|  } 
\eea 
The probability distribution over ribbon graphs is given by 
\bea 
W ( [ 1^n ] , r ) = { 1 \over n!  |\Aut ~ ( r ) | } 
\eea
Thus, the  leading asymptotics comes from a probability distribution over all bi-partite ribbon graphs, where each 
contributes an inverse of the order of its automorphism group. The contribution at order $1/n^2  $ in $ Z_3 ( n )$ comes from the $ [\gamma ] = [1^{ n-2} , 2 ] $. This contribution is associated with the probability distribution 
\bea 
W ( [1^{ n-2}, 2  ]  , r ) = { 1 \over ( n-2) ! 2  } { |\Aut ( r ) \cap [1^{ n-2} , 2 ] | \over |\Aut ~ ( r ) |  } 
\eea
As a generalization of this link to probability distributions, if we consider the coefficient of 
${ 1 \over ( n) ( n -1) \cdots ( n - k+1) }  $ in \eqref{informalZ3}, we have a sum  of symmetry factors over a finite set of partitions 
of the form $ p = [1^{ n-k} , q ] $ with $ q \vdash k ; q_1 =0$. The contribution of a given ribbon graph equivalence class (labelled by $r$) to 
this sum is proportional to a probability distribution over ribbon graphs. Let $\mathcal{S}_k$ be the set of partitions of this form specified $ p = [1^{ n-k} , q ]$.  For the subset $S_k$  there is a probability distribution 
\bea 
W ( \mathcal{S}_k , r ) = { 1 \over \sum_{ p \in \mathcal{S}_k  } \Sym ( p )  } \sum_{ p \in \mathcal{S}_k  } { |\Aut ( r ) \cap [p] | \over |\Aut ~ ( r ) |  } 
\eea 
It is interesting to describe the geometrical characteristics of the ribbon  graphs which lead to the largest contributions for each $p$. Since ribbon graphs also correspond to Belyi maps (see for example \cite{LandoZvonkin}), we may phrase  this question in terms of characteristics such as Galois invariants of Belyi maps. We leave these as interesting questions for the future.

\section{ Asymptotics in terms of  Stirling numbers and generalization to $ Z_{ 3; m} $ }
\label{sect:expStirl}

In this section we show that the asymptotic expansion of  $ Z_{ 3 } ( n ) $, which is the same as that of $ Z_{ 3 ; 1 } ( n )$, involves the well-known Stirling numbers of the second kind. This shows that integers obtained by \cite{VK} are expressible in terms of symmetry factors of restricted partitions multiplied by these Stirling numbers.  The
 same structure  holds  true for asymptotic expansions of  $ Z_{ 3 ; m } $ for higher $m$. 

\subsection{$Z_{ 3 } ( n ) \sim Z_{3 ; 1}(n)$ in terms of Stirling numbers }

We use the defining property of the Stirling numbers of the second kind $S ( k + r ,  k ) $ \cite{StirSec}
\bea 
{ 1 \over ( 1 - x ) ( 1 - 2x ) \cdots ( 1 - k x ) } = \sum_{ r = 0 }^{ \infty } S ( k + r  , k ) x^{ r  } 
\eea
with the substitutions $ x \rightarrow n^{-1} , k \rightarrow ( k-1)$ to obtain the large $n$ expansion  
\bea 
&& { 1 \over n ( n-1) \cdots ( n - k+1) } = 
{ 1 \over n^{ k } }  { 1 \over  ( 1- n^{-1} ) ( 1- 2 n^{ -1} ) \cdots ( 1 -( k - 1) n^{ -1} ) } \cr 
&& = \sum_{ r = 0}^{ \infty } S ( k- 1  + r , k -1 )  (n^{-1})^{ r + k  } 
\eea
We rewrite using  $K \le  n $ 
\bea 
\label{z31}
 && S_{ 3 ; 1 } ( n , K ) = n! \sum_{ k =0 }^K  { ( n -k)!\over n! }  \sum_{ q \vdash k  : q_1 = 0 } \Sym ( q ) \cr 
&& = n!   \Big(1+ \sum_{ k =2 }^K \sum_{ r = 0 }^{ \infty} S( k - 1 + r , k-1) ( n^{ -1} )^{ k + r  }  \sum_{ q \vdash k  : q_1 = 0 } \Sym ( q )     \Big)  \cr 
&& 
 = n!   \Big(1+ \sum_{ r = 0 }^{ \infty}     \sum_{ k = 2 }^{ K } ( n^{ -1} )^{ k + r  } S(k - 1+r , k-1)    \sum_{ q \vdash k  : q_1 = 0 } \Sym ( q )   \Big) \cr 
 &&  = n!   \Big(1+ \sum_{ l = 2 }^{ \infty}   ( n^{ -1} )^{l }   \sum_{ k = 2 }^{\min (K,l)} S(l - 1, k-1)    \sum_{ q \vdash k  : q_1 = 0 } \Sym ( q )   \Big) 
\eea
To understand the second line of the above, note that there no partitions $q$
of $k=1$ with  $q_1=0$.

Thus, we have the expansion 
\bea S_{ 3 ; 1; K } ( n )  =n!  \sum_{l=0}^{\infty} a_l (K) \, (n^{-1})^{l}
\eea 
where the coefficients $a_l(K)$ are given by 
\bea
\label{coeffs31}
&&
a_0(K)= 1 \,, \qquad a_1(K) = 0 \crcr
&&
a_l (K)=  \sum_{ k = 1 }^{\min (K,l)} S(l - 1, k-1)    \sum_{ q \vdash k  : q_1 = 0 } \Sym ( q )  
\qquad  l \ge 2
\eea

It is useful, as in \eqref{informalZ31} to express the result  \eqref{z31} for the  asymptotic 
$1/n$ expansion informally as 
\bea
 { Z_{ 3 ; 1 } ( n )\over n! }  &\sim&    \Big(1+ \sum_{ r = 0 }^{ \infty}     \sum_{ k = 2 }^{ \infty  }
 ( n^{ -1} )^{ k + r  } S(k - 1+r , k-1)    \sum_{ q \vdash k  : q_1 = 0 } \Sym ( q )   \Big) \cr 
& \sim &  \Big(1+ \sum_{ l = 2 }^{ \infty}   ( n^{ -1} )^{l }   \sum_{ k = 2 }^{l } S(l - 1, k-1)    \sum_{ q \vdash k  : q_1 = 0 } \Sym ( q )   \Big) 
\eea
It also follows from Theorem \ref{theosymp} and Theorem 
\ref{z3asympt}  that $ { Z_{ 3  } ( n )\over n! } $ has
the same asymptotics as  ${ Z_{ 3 ; 1 } ( n )\over n! } $ so that we have 
 \bea\label{resultStirling} 
 \boxed{
 { Z_{ 3  } ( n )\over n! }   \sim   \Big(1+ \sum_{ l = 2 }^{ \infty}   ( n^{ -1} )^{l }   \sum_{ k = 2 }^{l } S(l - 1, k-1)    \sum_{ q \vdash k  : q_1 = 0 } \Sym ( q )   \Big) 
 }
\eea

\

\noindent{\bf Examples.} We illustrate the above formulae and
check if the coefficient appears in the series \eqref{z31n}. 

-  For $l=2$, $K=n> 2$
 \bea
 &&
 a_2 =  \sum_{ k = 1 }^{2} S(1, k-1)    \sum_{ q \vdash k  : q_1 = 0 } \Sym ( q )  \crcr
 &&
 = S(1,0)  \sum_{ q \vdash 1  : q_1 = 0 } \Sym ( q )   + 
 S(1,1)  \sum_{ q \vdash 2  : q_1 = 0 } \Sym ( q ) 
  = S(1,1) \Sym([2]) = 2 
 \eea
 That agrees with \eqref{z31n}. 
 
- For $l=3$, $K=n> 3$
 \bea
 &&
 a_3 =  \sum_{ k = 1 }^{3} S(2, k-1)    \sum_{ q \vdash k  : q_1 = 0 } \Sym ( q )  \crcr
 &&
 = S(2,0)  \sum_{ q \vdash 1  : q_1 = 0 } \Sym ( q )   + 
 S(2,1)  \sum_{ q \vdash 2  : q_1 = 0 } \Sym ( q ) 
   + 
 S(2,2)  \sum_{ q \vdash 3  : q_1 = 0 } \Sym ( q ) \crcr
 && 
  = S(2,1)   \Sym([2])  + S(2,2)   \Sym([3]) 
  =   2 + 3 =5 
 \eea
that once again agrees with \eqref{z31n}. 

More generally, the formula matches with asymptotic expansion as given in OEIS 
A279819.

\

\subsection{$Z_{ 3 ; m }(n)$ in terms of Stirling numbers}
We  conjecture here  the asymptotic  series expansion for  $Z_{ 3 ; m }(n) $, for general finite $m$, 
according to similar arguments given above. 
We recall that 
\bea \label{Z3mn}
Z_{ 3 ; m } ( n )  = \sum_{ p \in  \cP_m ( n ) } \Sym ( p )  = \sum_{ \substack { 
p \vdash n \\ p_1 = \cdots = p_{m-1} = 0 } ;\;  p_m > 0  } \Sym ( p ) 
\eea
with $m\le  n$. 

For $m > 1$, there are in fact more constraints on the partition than $p \vdash n,$ $p_1 = \cdots = p_{m-1} = 0$, $p_m > 0$, in the above sum \eqref{Z3mn}. 
Indeed, consider the Euclidean division
\bea
n = m l_1 + l_2, \qquad    0 \le  l_2 <  m,   \qquad l_1 \ge 1
\eea
 Two cases should be discussed 
pertaining to the value of the remainder: 
either $l_2$ equals 0 or  does not. 

If $l_2 = 0$, then $n = ml_1$ and we claim the dominant
term in \eqref{Z3mn} is given by 
\bea 
p = [m^{l_1}] 
\eea
If $l_2 >0$, then  $ n -l_2 = ml_1$, then the following term should
be the dominant one: 
\bea
p = [m^{l_1 -1},  q ]\,, \qquad q \vdash m+l_2
\eea 
where $q$ should also obey  $q_1 = q_2 = \dots =q_{m} = 0$. 
A quick inspection shows the unique possibility  $q= [m+l_2]$, 
 hence $p = [m^{l_1 -1},  m+l_2 ]$. 

Depending on $l_2$, we use the notation 
$\delta_{l_2=0}  = 1$, if $l_2=0$ and $\delta_{l_2=0}  = 0$, otherwise,
and  $\delta_{l_2>0}  = 1$, if $l_2>0$ and $\delta_{l_2>0}  = 0$, otherwise.
We expand the partial sum 
\bea 
S_{ 3 ; m; K }(n) 
&=&
\delta_{l_2=0}  \; l_1  ! m^{ l_1  }  
 + \delta_{l_2>0} \;  ( l_1  -1 )! m^{  (l_1  - 1 ) } (m+l_2) \crcr
 &+&
  \sum_{ k =2}^K ( l_1  - k )! m^{  (l_1  - k ) }  \sum_{ q \vdash  m k +l_2 : q_1 = q_2 = \cdots = q_m = 0 } \Sym(q) 
\eea
where it is understood that $K\le l_1$, as  $n$ and, therefore, $l_1= (n-l_2)/m$ go to infinity. 

For $l_2= 0$, we can further expand $S_{ 3 ; m; K }(n)$ and obtain
the coefficients of the conjectured asymtotic expansion of $Z_{3;m}(n)$. 
In an analogous way to the steps  leading to \eqref{z31}, we 
introduce
\bea
\mathcal{F}_{m,k}(l_2) := \sum_{ q \vdash  m k +l_2   : q_1 = q_2 = \cdots = q_m = 0 } \Sym(q)
\eea
and write: 
\bea 
S_{ 3 ; m; K }(n) &=&
 l_1  ! m^{ l_1  }  \Big[1+ 
  \sum_{ k =2}^K m^{ - k } \frac{( l_1  - k )!  }{ l_1  !} \;  \mathcal{F}_{m,k} (0)  \Big] \\
  &= &
  l_1  !  m^{ l_1 }   \Big[ 1+ 
  \sum_{ k =2}^K m^{  - k }   
   \sum_{ r = 0 }^{ \infty} S( k - 1 + r , k-1) ( l_1 ^{ -1} )^{ k + r  }  \;  \mathcal{F}_{m,k} (0)    \Big] \cr\cr
  &= &
  l_1  !  m^{ l_1 }   \Big[ 1+ 
 \sum_{ r = 2 }^{ \infty}  ( l_1 ^{ -1} )^{ r  }  \sum_{ k =2}^{\min (K, r)}  m^{ - k }   
   S( r - 1  , k-1)   \;  \mathcal{F}_{m,k}  (0)    \Big]
   \nonumber
\eea 
Thus, the coefficients of the expansion of  $S_{ 3 ; m; K }(n)/(l_1! m^{l_1})$ read off
\bea\label{coeffZ3m}
&& a_0(m,K)  = 1 \crcr
&& a_1(m,K)  = 0 \crcr
&& a_r(m, K)  = 
  \sum_{ k =2}^{\min (K, r)}  m^{r - k }   
   S(  r- 1 , k-1)  \;  \Sym_{m,k}    (0) 
\eea 
where in the last line $ r \ge 2$. 
This covers the case $m=1$ in \eqref{coeffs31}. 
Our conjecture is expressed as 
\bea \label{conjectureZ3m}
 { Z_{ 3 ; m } ( n )\over {(\frac{n}{m})! m^{ \frac{n}{m}} } }  \sim    
 1+ 
 \sum_{ r = 2 }^{ \infty}  (n ^{ -1} )^{ r  }  \sum_{ k =2}^{\min (K, r)}  m^{ r- k }   
   S( r - 1  , k-1)   \;  \Sym_{m,k}  (0)   
\eea
Applying \eqref{coeffZ3m} to $m=2$,  the procedure computing the 
coefficients of  the expansion of $S_{3,2}(n)/(2^{ n/2 } ( n/2 ) !)$ yield at $n=50$
\bea
&&
r= 2  , \qquad  4  \crcr
&&
r= 3    , \qquad   32 \crcr
&&
r= 4   , \qquad    215  \crcr
&&
r= 5   , \qquad    1541  \crcr
&&
r= 6   , \qquad    14658  \crcr
&&
r= 7   , \qquad    180246 \crcr
&& 
r= 8    , \qquad   2425061  \crcr
&&
r= 9   , \qquad    33315155  \crcr
&&
r= 10   , \quad\;    478703544
\eea
a sequence unlisted in OEIS. 

We now address the case $l_2>0$ and  express $S_{ 3 ; m; K }(n)$ as 
\bea 
&&
S_{ 3 ; m; K }(n) =
 (l_1-1)  ! m^{ l_1-1  }  \Big[ (m+l_2) + 
  \sum_{ k =2}^K m^{ - (k-1) } \frac{( l_1  -1 -( k-1) )!  }{ (l_1 -1) !}   \;  \mathcal{F}_{m,k}    (l_2 ) \Big] \cr\cr
  & & = 
 (l_1-1)  ! m^{ l_1-1  }    \Big[ (m+l_2)+ 
  \sum_{ k =2}^K m^{ - (k-1) } 
   \sum_{ r = 0 }^{ \infty} S( k - 2 + r , k-2) ( l_1 ^{ -1} )^{ k -1+ r  } 
 \;  \mathcal{F}_{m,k}    (l_2 )  \Big] \cr\cr
  & & =  
 (l_1-1)  ! m^{ l_1-1  }     \Big[ (m+l_2)+  
 \sum_{ r = 2 }^{ \infty}  ( l_1 ^{ -1} )^{ r -1 }  \sum_{ k =2}^{\min (K, r)}  m^{ - (k-1) }
   S( r - 2  , k-2)   \;  \mathcal{F}_{m,k}    (l_2 )   \Big]
   \cr\cr
  & & =  
 (l_1-1)  ! m^{ l_1-1  }      \crcr
 && \times   \Big[ (m+l_2)+  
 \sum_{ r = 1 }^{ \infty}  (n-l_2)^{ -r}  \sum_{ k =2}^{\min (K, r+1)}  m^{ r- (k-1) }
   S( r - 1  , k-2)   \;  \mathcal{F}_{m,k}    (l_2 )   \Big]   
    \cr\cr
  & & =  
 (l_1-1)  ! m^{ l_1-1  }    \crcr
 && \times  \Big[ (m+l_2)+  
 \sum_{ r = 1 }^{ \infty}  n^{-r}\Big(\sum_{i=0}^{\infty} C_{r,i} (\frac{l_2}{n})^i \Big) \sum_{ k =2}^{\min (K, r+1)}  m^{ r- (k-1) }
   S( r - 1  , k-2)   \;  \mathcal{F}_{m,k}    (l_2 )   \Big]
   \nonumber
\eea 
where we used the notation $C_{r,i} =  \frac{r(r-1)(r-2)\dots (r-i-1)}{i!}$ for the generalized binomial coefficient. We obtain
\bea 
&&
    S_{ 3 ; m; K }(n) 
=
 (l_1-1)  ! m^{ l_1-1  }    \crcr
 && \times  \Big[ (m+l_2)+  
 \sum_{ r = 1 }^{ \infty}\sum_{i=0}^{\infty} C_{r,i}  \, \frac{ l_2^{i}  }{n^{r+i}}   \sum_{ k =2}^{\min (K, r+1)}  m^{ r- (k-1) }
   S( r - 1  , k-2)   \;  \mathcal{F}_{m,k}    (l_2 )   \Big]
    \cr\cr
 &&
=
 (l_1-1)  ! m^{ l_1-1  }    \crcr
 && \times  \Big[ (m+l_2)+  
 \sum_{ s = 1 }^{ \infty }  n^{-s}   \sum_{r=0}^{s} l_2^{s-r}   \, C_{r,s-r}     \sum_{ k =2}^{\min (K, r+1)}  m^{ r- (k-1) }
   S( r - 1  , k-2)   \;  \mathcal{F}_{m,k}    (l_2 )   \Big]
   \nonumber
\eea 

The coefficients of the expansion of  $S_{ 3 ; m; K }(n)/( (l_1-1)  ! m^{ l_1-1  } (m+l_2))$
are given by,
\bea
&& a_0(m,K)  = 1 \\
&& a_s(m, K)  =  \frac{1}{(m+l_2)}
 \sum_{r=0}^{s} l_2^{s-r}   \, C_{r,s-r} 
  \sum_{ k =2}^{\min (K, r+1)}  m^{ r+1 - k }   
   S( r - 1  , k-2)  \;   \mathcal{F}_{m,k}    (l_2 ) 
      \nonumber
\eea 
where the last line holds for $s\ge 1$. 

In this case, $l_2 >0$, we conjecture the  asymptotics, 
 \bea\label{conjectureZ3ml2}
&&
 { Z_{ 3 ; m } ( n )\over {(\frac{n}{m}-1)! m^{ \frac{n}{m} -1} } (m+l_2)  }  \sim    
 \cr\cr 
 &&
 1+ 
 \sum_{ s = 1 }^{ \infty }  n^{-s}   \sum_{r=0}^{s} l_2^{s-r}   \, C_{r,s-r}     \sum_{ k =2}^{\min (K, r+1)}  m^{ r- (k-1) }
   S( r - 1  , k-2)   \;   \mathcal{F}_{m,k}    (l_2 )  
\eea

\subsection{Discussion: Non-perturbative asymptotics of $ Z_{ 3 } (n ) $ }\label{sec:NPA}

We have established the large $n$ 
asymptotic series for $ Z_3 ,  Z_{3; 1 } $, both of which have the same large $n$ series. We have argued for and  conjectured the large $n$ series for $Z_{3, m }$, for any $m  \sim \cO ( 1 ) $ as $ n \rightarrow \infty $. Drawing on analogies with non-perturbative expansions in QFT and quantum mechanics, it is natural to ask whether  our knowledge of the asymptotic expansions of  $ Z_{ 3 ; m } $ can be collected into a non-perturbative expansion for $ Z_3$. Although $ Z_{ 3 } = \sum_{ m } Z_{ 3 ;   m } $ this is not straightforward since the number of terms in the sum over $m$ goes to infinity as $ n  \rightarrow \infty $.

Let us first explain the analogy in more detail.  Instanton expansions in QFT where QFT observables 
are expressed as an approximation of the form 
\bea\label{instantons}  
F ( g ) \sim  S_0 ( g ) + e^{ - 1 \over g } S_1 ( g ) + e^{ - 2 \over g } S_2 ( g ) + \cdots 
\eea
Here $ S_0 ( g ) , S_1 ( g )  , \cdots $ are power series in powers of 
 $ g$,  the coupling constant, which are asymptotic expansions in the limit $g \rightarrow 0$. The successively higher instanton numbers are exponentially suppressed in the limit. See \cite{Marino} for a review of this subject. 

Based on the analogy, we can ask if it is possible to make sense of an expansion of the form 
\bea\label{Z3complete}  
{ Z_{ 3 } \over n! }  \sim  S_{ 3 ; 1 } + f_2 ( n ) S_{ 3 ; 2 } +  f_3 ( n ) S_{ 3 ; 3 } + \cdots   ~~~ ? 
\eea
$ S_{ 3 ; 1 } $ is a power series analogous to $ S_0 ( g)$, the perturbative term in QFT. $ S_{ 3 ; 2 },S_{ 3 ; 3 } \cdots $ are likewise power series in $ { 1 \over n }$ analogous to $ S_1 ( g ), S_2 ( g ) \cdots $ in QFT. 
$f_2 (n)$ is super-exponentially suppressed compared to $1$, $ f_3 ( n )$ is super-exponentially suppressed compared to $ f_2 (n)$ etc. 
As we saw, the series  $S_{ 3 ; 1 } $ is obtained from $ { Z_{ 3 ; 1} \over n! }$ where $Z_{ 3 ; 1 }   $ \eqref{Z3mn} is the sum of $  \Sym (p) $ for partitions where the minimum part has length $1$. Finite truncations 
$ S_{  3 ; 1 ; K } $ are obtained by  taking $ n \gg  K $ and summing terms in $ { Z_{ 3 ; 1  } \over n! } $ 
with $ n - K $ parts of length $1$. 

 The term 
$f_2 ( n ) S_{ 3 ; 2 } $ is obtained from $ Z_{ 3 ; 2 }/n!  $ which is the sum of $ \Sym (p) $ for partitions that have no cycles of length $1$ and a non-zero number of parts of length $2$. Finite truncations $ f_2 ( n ) S_{ 3  ; 2 ; K } $ are obtained by taking $ n\gg  K $ and considering terms in $ Z_{ 3 ; 2 }$  with at least $( n - K) $ parts of length $2$: 
 \bea 
 f_2 ( n ) = { (\lfloor n/2 \rfloor )! 2^{\lfloor  n/2 \rfloor} \over n ! } 
 \eea
 We see that $ f_2 ( n )$ is super-exponentially suppressed compared to $ f_1 ( n ) = 1 $. 
 Similarly $ f_3 ( n ) S_{ 3 ; 3  } $ is obtained from $ Z_{ 3 ; 3}/n! $ which is the sum of $ \Sym (p )$ for partitions 
 having no cycles of length $1,2$ and a minimum part of length $3$. $ f_3 ( n )$ is the symmetry factor for a 
 partition with the largest number of $3$, divided by $ n! $. 
 \bea 
 f_3 ( n )  = { (\lfloor n/3 \rfloor)! 3^{\lfloor  n/3 \rfloor} \over n ! } 
 \eea 
  The series $ S_{ 3 , 3 } $ starts with $1$ and any finite order truncation 
 $ S_{ 3 ; 3 ; K  } $ is obtained by taking $ n \gg K $ and summing $ \Sym (p)$ over partitions having
  $\lfloor n /3 \rfloor - K $ parts of length $3$. It is easy to see that $ f_3 ( n ) $ is super-exponentially suppressed compared to $ f_1 ( n )$ 
 and $ f_2 ( n )$. 
 
One approach to making sense of \eqref{Z3complete} is to interpret it as 
a sequence of asymptotic expansions related to $ Z_{ 3 } ( n )$ 
\bea 
Z_3 ( n ) \sim S_{ 3 , 1  } ( n ) 
\eea
After subtracting $ Z_{ 3 , 1  } ( n )$ we have 
\bea 
{ Z_3 ( n ) - Z_{ 3,1} (n ) \over n! f_2 ( n ) }  \sim S_{ 3 , 2 } ( n) 
\eea
After further subtracting $ Z_{ 3,2} (n)$ we have 
\bea 
{ Z_3 ( n ) - Z_{ 3,1} (n ) - Z_{ 3,2} (n)  \over n! f_3 ( n ) }  \sim S_{ 3,3} (n )  
\eea
Indeed for any finite $m$, 
\bea 
{ Z_3 ( n ) - Z_{ 3,1} (n ) - \cdots - Z_{ 3 , m } ( n )    \over n! f_{ m+1}  ( n ) } 
\sim S_{ 3 , m+1 } ( n ) 
\eea
 $ Z_{ 3 } ( n )/n! $ is not unique in having the asymptotic expansion $S_{3,1} (n) $ :  $ ( Z_{ 3} ( n ) - Z_{ 3 ; 2 } )/n! $ has the same expansion, it is tempting to conjecture that the above equations uniquely 
 determine $Z_3(n)$. Thus we present  a conjecture.

\noindent{\bf Conjecture.}  If a function $ F (n ) $ obeys the properties 
\bea 
F ( n )  & \sim &  S_{ 3 , 1  } ( n ) \cr 
{ F ( n ) - Z_{ 3,1} (n ) \over n! f_2 ( n ) }  & \sim &  S_{ 3 , 2 } ( n) \cr 
{ F  ( n ) - Z_{ 3,1} (n ) - Z_{ 3,2} (n)  \over n! f_3 ( n ) }  & \sim  & S_{ 3,3} (n )  \cr 
& \vdots &  \cr 
{ F  ( n ) - Z_{ 3,1} (n ) - \cdots - Z_{ 3 , m } ( n )    \over n! f_{ m+1}  ( n ) } 
& \sim &  S_{ 3 , m+1 } ( n ) 
\eea
for all finite $m$, then $ F ( n ) = Z_3 ( n ) $.

\section{Higher rank tensors} 
\label{sect:hr}

There exists an enumeration formula for 
higher rank $d$ tensor model observables   in terms of sums of powers of symmetry factors \cite{JoSan1}.  It is then natural to ask, in full generality, the question of the asymptotic expansion of that counting.
At this point,  from the asymptotic dominance
of small parts, and since one easily realizes that this should hold independently 
of the rank of the tensor invariant, we conjecture below the series expansion of that counting of rank $d$ tensor.

For rank $d $ tensors, 
$Z_{ d  }(  n) $ counts the number of rank $d$ tensor invariants.  We have in a similar way as above 
\bea
Z_{ d  }(  n ) =   \sum_{ p \vdash n }  (  \Sym ~ p )^{ d-2 } 
=  \sum_{ m = 1 }^{ \lfloor  \frac{n}{2} \rfloor }  Z_{ d ; m  } ( n )  + Z_{d, n} (n) 
\eea
with,  for $1 \le m \le \lfloor  \frac{n}{2} \rfloor $. 

For the next developments, we assume for simplicity that $n = m l$, i.e. that
$n$ is a multiple of $m$. The generic case should require a bit more work. 

We introduce the partial sum, assuming that $n$ and $l$ are large enough, 
\bea
S_{ d ; m; K  } (n)
&=& 
\sum_{ k =0 }^{K} \sum_{ q \vdash  m k  : q_1 = q_2 = \cdots = q_m = 0 } 
\Big( \Sym (  [ m^{ l   - k } ,  q ] )  \Big)^{ d-2 }  \,, \qquad 
1 \le m \le \lfloor  \frac{n}{2} \rfloor 
 \crcr
S_{ d ; n; K  } (n)   &=& \Big(  \Sym([n])  \Big)^{ d-2 }  =  n ^{ d-2 }  = Z_{d, n} (n) 
\eea 
By similar techniques previously introduced, we can work out the following expansions: 
\bea
S_{ d ; m; K  } (n) &=& 
\sum_{ k =0 }^{ K  } \sum_{ q \vdash  m k  : q_1 = q_2 = \cdots = q_m = 0 } 
\Big(   m^{ l  - k }  ( l  - k)!  \;   \Sym ( q  )  \Big)^{ d-2 } 
 \crcr
 &=&
\sum_{ k =0 }^{ \min (K,l)  } 
\Big(   m^{ l  - k }  ( l - k)!\Big) ^{d - 2}
\sum_{ q \vdash  m k  : q_1 = q_2 = \cdots = q_m = 0 } 
\Big( \Sym ( q  )  \Big)^{ d-2 }  
\eea
When $n \to \infty$,  the coefficient $\Big(   m^{ l - k }  ( l  - k)!\Big) ^{d - 2}$ becomes less and less dominant as soon as $m>1$. 

Let us restrict to $m=1$, and conjecture an asymptotic expansion
of that sector  (we conjecture to be the dominant order is given by the 
fixed order $m=1$): 
\bea
S_{ d ; 1; K  } (n)
&=& 
\sum_{ k =0 }^{ K } \sum_{ q \vdash  k  : q_1 = 0 } 
\Big( \Sym (  [ 1^{ n- k } ,  q ] )  \Big)^{ d-2 } 
= 
\sum_{ k =0 }^{ K } \sum_{ q \vdash  k  : q_1 = 0 } 
\Big( (n-k)!  \; \Sym ( q )  \Big)^{ d-2 } \crcr
& =& (n ! ) ^{d-2}\Bigg[ 1 + 
\sum_{ k =2 }^{ K }  \Big( \frac{ (n-k)! }{ n! } \Big)^{ d-2 } \sum_{ q \vdash  k  : q_1 = 0 } 
\Big( \Sym ( q )  \Big)^{ d-2 }  \Bigg]  \crcr
& =& (n ! ) ^{d-2}\Bigg[ 1 + 
\sum_{ k =2 }^{ K }  \Big(\sum_{ r = 0 }^{ \infty} S( k - 1 + r , k-1) ( n^{ -1} )^{ k + r  }  \Big)^{ d-2 } \sum_{ q \vdash  k  : q_1 = 0 } 
\Big( \Sym ( q )  \Big)^{ d-2 }  \Bigg]\crcr
&=& 
 (n ! ) ^{d-2} \Bigg[ 1 +
\sum_{ k =2 }^{ K }  \;  \sum_{p=0}^\infty  n^{  - p  - (d-2)k } \crcr
 && \times  \; \sum_{  \sum_{i=1}^{d-2} r_i = p }  \; \Big(\prod_{i=1}^{d-2} S (  k-1 +  r_i  , k-1 )   \Big)   \sum_{ q \vdash  k  : q_1 = 0 } 
\Big( \Sym ( q )  \Big)^{ d-2 }\Bigg] 
\eea 
Using a change of variables $ \tilde p = p+(d-2)k$ (and rename $\tilde p \to p$), and $ \tilde r_i = r_i +k$  (and rename $\tilde r_i \to r_i$), then swapping the two sums over $k$ and $p$, we obtain an expression generalizing \eqref{z31}, 
\bea
S_{ d ; 1; K  } (n)
 &=& 
 (n ! ) ^{d-2} \Bigg[ 1 +
 \;  \sum_{p=2(d-2)}^\infty  n^{  - p } 
 \\ 
 & \times &   \sum_{ k =2 }^{ \min (K, \frac{p}{d-2} ) } 
 \sum_{  \sum_{i=1}^{d-2} r_i = p  ; \; \text{and} \; r_i \ge k }  \; \Big(\prod_{i=1}^{d-2} S (  r_i - 1 , k-1 )   \Big)   \sum_{ q \vdash  k  : q_1 = 0 } 
\Big( \Sym ( q )  \Big)^{ d-2 }\Bigg] 
\nonumber 
\eea 
Thus we conjecture that  $\frac{Z_{ d ; 1  }}{  (n ! ) ^{d-2} }  \sim    \sum_{p=0}^\infty A_{d; p} / n^{  p  }   $  has the coefficients 
\bea
&&
A_{d:0} =  (n ! ) ^{d-2} \cr\cr
&&
A_{d:p} 
=\sum_{ k =2 }^{ \min (K, \frac{p}{d-2} ) } 
 \sum_{  \sum_{i=1}^{d-2} r_i = p  ; \; \text{and} \; r_i \ge k  }  \; \Big(\prod_{i=1}^{d-2} S (  r_i - 1 , k-1 )   \Big)   \sum_{ q \vdash  k  : q_1 = 0 } 
\Big( \Sym ( q )  \Big)^{ d-2 }
\eea
for $p \ge 2(d-2)$. 
The same asymptotic series should hold for $ \frac{Z_{ d  }}{  (n ! ) ^{d-2} }$.
Restricted to $d=3$, we recover \eqref{coeffs31} as expected.

\section{Conclusion}
\label{ccl}

We have determined the asymptotic expansion of the
counting of rank 3 (unitary) tensor invariants. We have exploited the counting formula 
in terms of a sum of symmetry factors of partitions. A general principle we have found useful is 
that these sums of symmetry factors are dominated by partitions with a large multiplicity of a small part. 
The asymptotic series has been provided and its coefficients determined at all
orders : the key results are \eqref{informalZ31}\eqref{resultStirling}.  
 As an interesting feature, we express these coefficients
as a sum of Stirling numbers of the second kind. 
We also conjecture similar formulae 
for the enumeration of any rank $d$ tensor invariants, and expect similar proofs will work. 
The same general principle allows formulae for sums of symmetry factors of restricted partitions, which we have denoted $Z_{ 3 ; m }$.

It would be interesting to investigate
the asymptotics of  connected tensor invariants. The connected invariants are obtained from the disconnected ones, $ Z_3 (n ) $ (for which we have derived the asymptotics),  by taking a plethystic logarithm \cite{JoSan1}. The sequence of connected invariants is known to high orders \cite{OEIS-connected}. 
 By inserting the asymptotic expansion of $Z_3(n)$  into the plethystic logarithm (PLOG) function, it should  be possible to 
 obtain  the asymptotic expansion of \cite{OEIS-connected}. 
Finally, we may ask if the same ideas developed in this work could be applied
to the analysis  of orthogonal tensor invariants \cite{Avohou:2019qrl}. 
That series  would be slightly different but we expect that the main principle
 discovered  in this work, i.e. partitions with a large multiplicity of a  small part will  dominate sums over symmetry factors,  would apply again in that situation. This deserves to be addressed thoroughly. 
 
 It is  instructive, in  the context of holography and brane physics,  to compare the asymptotics of tensor model counting with that of multi-matrix models, and to develop interpretations of the asymptotic results  \eqref{informalZ31}\eqref{resultStirling} in these contexts.  These asymptotic results have implications for the thermodynamics of quantum mechanical models based on tensor or multi-matrix models respectively. The multi-matrix case has been discussed in the context of AdS5/CFT4 and related gauge theories in \cite{AMMPV04} and more recently in \cite{Ber1806,RWZ,KZ-2021-05,KW2005}. The asymptotic counting we have done in this paper holds at large $N$. We are considering large $n$ invariants when $ N \gg n$. The super-exponential growth of $ Z_{ 3 } ( n )$ has the consequence of a vanishing Hagedorn temperature in this large $n$ limit \cite{Tseytlin}. It will be very interesting to investigate the fate of this Hagedorn behaviour in the finite $N$ tensor systems. The analogous investigation has been investigated for multi-matrix models in \cite{Ber1806,KW2005}. The role of tensor models in connection with M5-branes has been discussed in \cite{Tseytlin}. Beyond the counting of observables it is also interesting to look at the asymptotic behaviour of correlators, for example with motivations from quantum  information theoretic aspects of holography  \cite{Milekhin2008}.

\begin{center} 
{\bf Acknowledgements}
\end{center} 
SR is supported by the STFC consolidated grant ST/P000754/1 `` String Theory, Gauge Theory \& Duality” and  a Visiting Professorship at the University of the Witwatersrand, funded by a Simons Foundation grant (509116)  awarded to the Mandelstam Institute for Theoretical Physics. We thank  Fabien Vignes-Tourneret and Vaclav Kostosevec for discussions which led to the initiation of  this project. We also thank George Barnes, Robert de Mello Koch and Adrian Padellaro  for  interesting discussions on the subject of the paper.

\section*{Appendix}

\appendix

\renewcommand{\theequation}{\Alph{section}.\arabic{equation}}
\setcounter{equation}{0}

 \section{Proof:  of Lemma \ref{symq}}
 \label{app:proofLemsymq}
 
In this appendix, the proof of Lemma \ref{symq} is given. It divides into several 
cases that must be carefully checked. 

 \subsection{Case $k$ even: Proof of $\Sym[2^{k/2}] \ge  \Sym(q)$, $q \vdash k$, $q_1= 0$}
 
We want to  prove that $\Sym[2^{k/2}] > \Sym(q)$, for $q \vdash k$, with $q_1= 0$, 
for $k$ sufficiently large. We proceed by induction on $k$. 

\

Assume $k=2$. It is easy to see that   $\Sym[2^{k/2}] \ge  \Sym(q)$ for  $q\vdash 2$, $q_1=0$. The condition $ q_1=0$ means that $q=[2]$, so that  $\Sym(q) = 2 = \Sym[2^{2/2}] $. Let us assume this to be true for all  even $k'$ up to $k$, that is 
\bea
 \Sym[2^{k'/2}]  \ge  \Sym(q)  ,    \qquad 
q \vdash k', \qquad q_ 1= 0 \;, \hbox{ for $k'$ even and }~ k' \le k   
\eea 
We now prove for $k+2$ that $\Sym[2^{k/2 + 1}] \ge \Sym(q)$, $q \vdash k+2$, 
$q_1=0$. 
 
 Consider $q\vdash k+2$, we decompose $q= [2^m, q']$, with $q' \vdash k+2 - 2m$,
 $q'_2 = 0$ (no part of size 2 in $q'$). 
 
 \

\noindent{\bf  Case $0<m \le (k+2)/2$:} 
Then $q'  \vdash k+2 - 2m  =  k - 2(m-1) \le k$, so as $k - 2(m-1) $ is even then 
we write using our induction hypothesis 
\bea
&&
\Sym(q) =  \Sym [2^m, q'] = 2^m m! \, 
\Sym(q') \le 2^m m! \,   \Sym([2^{(k - 2(m-1))/2 }]) \crcr
&& 
 =  2^m m! \,    2^{(k - 2*(m-1))/2 } ( \frac{k - 2(m-1)}{2} )! \crcr
 &&
 =   2^{(k +2)/2 } m! \,    ( \frac{k - 2(m-1)}{2} )!  \le 2^{(k +2)/2 }  ( \frac{k + 2}{2} )!  = \Sym[ 2^{(k +2)/2 } ]
\eea

\noindent{\bf  Case $m =0$:} Let $l_0 >  2$ be the minimum part such that
$q_{l_0} > 0$, thus $q = [l_0^{q_{l_0}}, q']$ with 
$q'$  only containing parts of size strictly larger than $l_0$; $k+2 = \sum_{l \geq l_0} l q_{l}$.  
Then 
\bea
 \sum_{l >  l_0} l q_{l}  = k+2 - l_0 q_{l_0} < k+2 - 2 q_{l_0}  = k - 2( q_{l_0} -1) \le k 
\eea
 Then
we write using the induction hypothesis on $q'$
\bea
\Sym[l_0^{q_{l_0}}, q'] = l_0^{q_{l_0}} q_{l_0} ! \, 
\Sym[q'] \le  l_0^{q_{l_0}} q_{l_0} !  \,  \Sym[2^{ (k - 2( q_{l_0} -1))/2 }]
\eea
We want to show that
\bea
 l_0^{q_{l_0}} q_{l_0} !  \,  \Sym[2^{ (k - 2( q_{l_0} -1))/2 }] \le \Sym[2^{(k+2)/2}] \, . 
\eea
We evaluate the ratio
\bea
&&
\frac{ 2^{(k+2)/2} ((k+2)/2) ! }{     l_0^{q_{l_0}} q_{l_0} ! \, 2^{ (k - 2( q_{l_0} -1))/2 }
  ( (k - 2( q_{l_0} -1))/2)!  } = 
\frac{ 2^{(k+2)/2  - (k - 2( q_{l_0} -1))/2 } ((k+2)/2) ! }{     l_0^{q_{l_0}} q_{l_0} ! \,
  ( (k - 2( q_{l_0} -1))/2)!  }  \crcr
  && 
  = 
   \frac{ 2^{q_{l_0}  } ((k+2)/2) ((k+2)/2 - 1) ((k+2)/2 -2) \dots ((k+2)/2 - q_{l_0} +1) }{     l_0^{q_{l_0}} q_{l_0} !  } \crcr
  &&   = \frac{   \prod_{i=0}^{ q_{l_0} -1} (k+2 - 2i)      }{ l_0^{q_{l_0}} q_{l_0} !    }
   = \frac{   \prod_{i=0}^{ q_{l_0}  -1} (k+2 - 2i)      }{ \prod_{i=0}^{ q_{l_0} -1} ( l_0  q_{l_0}    - l_0 i )    }
    \ge  
    \frac{   \prod_{i=0}^{q_{l_0} -1} (k+2 - 2i)      }{ \prod_{i=0}^{q_{l_0}-1} ( l_0  q_{l_0}    -  2 i )    }
\eea
Using $k+2 \ge l_0 q_{l_0}$ and  $ (k+2 - 2i) \ge ( l_0  q_{l_0}    -  2 i ) $, for all $i=0,...,q_{l_0}-1$, the proof is completed.

 \subsection{Case $k\ge 11$ odd: Proof of $\Sym[3, 2^{(k-3)/2}] \ge  \Sym(q)$, $q \vdash k$, $q_1= 0$}
 
 We proceed again by induction on $k$. 
 
Let $k= 11$, the list of partitions of $q \vdash k$ with  parts  $\ge 2$ and their corresponding $\Sym(q)$ 
 are given by 
 \bea
 &&
\Sym [11]  = 11 , \qquad 
\Sym [9, 2] = 18 , \qquad  
\Sym [8, 3] =24 , \qquad 
\Sym [7, 4] =28 ,\crcr
&&  \Sym  [7, 2^2]=  56 , \qquad 
\Sym[6, 5] =30  , \qquad 
\Sym [6, 3, 2] =36 , \qquad 
\Sym [5, 4, 2] =40 , \crcr
&&
\Sym [5, 3^2] =90 ,\qquad 
\Sym [5, 2^3]= 240  , \qquad 
\Sym [4^2, 3] = 96 , \qquad 
\Sym [4, 3, 2^2] = 96 , \crcr
&&
\Sym [3^3, 2] = 324   , \qquad 
\Sym [3, 2^4]  = 1152\;. 
 \eea
 Hence $\Sym[3, 2^{(11-3)/2}] \ge \Sym (q)$, for any other $q\vdash k$ in the  above  list. Let us assume that the statement is true at order $k\ge k' \ge 11$
 \bea
 \Sym[3, 2^{(k'-3)/2}]  \ge  \Sym(q) ,    \qquad 
q \vdash k', \qquad q_ 1= 0 \;. 
\eea 
 Let us prove it at order $k+2$. 
 
 Consider the partition $q= [2^{q_2}, 3^{q_3}, q'] \vdash k+2 $, where 
 $q' \vdash  k+2 - 2q_2 -3q_3$. Either $q'$ is empty in 
 which case $k+2 - 2q_2 -3q_3=0$, 
 or $q'$ is a non empty partition with parts of size 4 or greater, 
 in which case $k+2 - 2q_2 -3q_3\ge  4$.

\
 
 \noindent{\bf Case $k+2 - 2q_2 -3q_3 = 0$.} Then $q'$ is an empty partition. Then 
 \bea
 k+2 =  2q_2 + 3q_3  \;, 
 \eea
 and we should compare $ \Sym (q) = 2^{q_2} q_2 ! 3^{q_3} q_3 !  $   (for $ q_3 >1$)
and $\Sym[3,2^{(k+2-3)/2}]$. 
 If $k+2 \ge 13$, and is odd, then $q_3 > 1$ and should be an
 odd number. 
 From now, $q_3 \ge 3$.

 We write
 the ratio (with $k+2 -3 = k-1\ge 10$)
 \bea
 \frac{\Sym[3,2^{(k+2-3)/2}] }{ \Sym(q) } 
 & =& 
   \frac{ 3 \cdot  2^{ q_2  + 3(q_3 -1)/2}   ((  2q_2 + 3q_3 -3)/2 ) !  }{2^{q_2} q_2 ! 3^{q_3} q_3 ! } \crcr
 & = &  \frac{ 3 \cdot  2^{  q_3 + (q_3 -3)/2}   ( q_2  + q_3+ (q_3 -3)/2 ) !  }{ 3^{q_3} q_2 !  q_3 ! } 
 \eea
 Assuming $q_3 = 3$, then $k+2 = 2q_2 +9$, so $k-7= 2q_2 \ge 4$, $q_2 \ge 2$, such that
 \bea
  \frac{\Sym[3,2^{(k+2-3)/2}] }{ \Sym[q] } =
   \frac{ 3 \cdot  2^{  3 }   ( q_2  + 3 ) !  }{ q_2 !  3^{3}  3 ! } 
   =   \frac{ 8 \cdot  ( q_2  + 3 )( q_2  + 2 )( q_2  + 1)  }{ 9 \cdot 6  } 
 \eea 
 Note that $q_2 = 0 $ would compromise this result. Indeed, this is what
 is happening for $k=9$ such that $\Sym [3^3] \ge \Sym[3,2^3]$. 
 Our condition $k \ge 11$ ensures that this does not happen. 
 
Now,  assume $q_3>3$. As $q_3$ is odd, we must have $q_3 \ge 5$, 
so that $(q_3-3)/2 - 1 \ge 0$. Then 
 \bea
 && 
  \frac{\Sym[3,2^{(k+2-3)/2}] }{ \Sym[q] } 
 \ge 
   \frac{    2^{  (3q_3 -1)/2 } \prod_{i=0}^{(q_3 -3)/2 -1 } ( q_2  + q_3 +  (q_3 -3)/2  -i )    }{ 3^{q_3-1}  }    \frac{( q_2 + q_3 ) !   }{   q_2 !  q_3 ! } \cr\cr
 && \ge 
   \frac{    2^{  (3q_3 -1)/2 }  }{ 3^{(q_3-1)}  }    \prod_{i=0}^{(q_3 -3)/2 -1 } ( q_2  + q_3 +  (q_3 -3)/2  -i )      \cr\cr
 && \ge 
   \frac{  2^{  (3q_3 -1)/2 }  }{ 3^{(q_3-1)}  }   \frac{   \prod_{i=0}^{(q_3 -3)/2 -1 } ( q_2  + q_3  +1 )    }{ 3^{(q_3-1)/2}  }  
   \ge   
   \frac{  2^{  (3q_3 -1)/2 }     4 ^{(q_3 -3)/2 -1}  }{ 3^{q_3-1}  } 
   =     
     \frac{  2^{  (3q_3 -1)/2  + q_3 - 3 -2}   }{ 3^{q_3-1}  }   \cr\cr
   &&
   \ge 
        \frac{  2^{  (5q_3 -11)/2 }   }{ 3^{q_3-1}  } 
\eea
Checking the exponent, we get for any $q\ge 5$, 
\bea
&&
(5q - 11 ) \ln 2  -   2(q -1) \ln 3  
 =  -  11 \ln  2   + 2 \ln 3  + q ( 5 \ln 2   - 2 \ln 3   ) \crcr
 && = 
 (q-5)( \ln 2^5   -  \ln 3^2  ) + 
  \ln 2^{14} - \ln 3^8 
 \ge 0 
\eea
and this ends the proof of the current case.

\ 

 \noindent{\bf Case $k+2 - 2q_2 -3q_3 \ge 4$.} 
 In this case, $q' \vdash k+2 - 2 q_2 -3q_3$ is non empty. 
 
 - Subcase 1: $q_2=q_3=0$:  Consider the smallest part $l_0 \ge 4$, 
 such that $ q_{l_0} >0$,  and write 
  $k+2 = l_0 q_{l_0} + \sum_{l>l_0} l q_l$. Then, 
 $k+2 - l_0 q_{l_0}  \le k +2 - 4q_{l_0} \le k - 2 \le k $. Define 
 $q'' \vdash k+2 - l_0 q_{l_0} $, such that
 $q' = [ l_0^{ q_{l_0} }, q'']$ where 
 $q'' \vdash k+2 - l_0 q_{l_0} \le k$ and the smallest part of $q''$ 
 is of size $\ge l_0+1$.

a) If $l_0$ is even, then $k+2 - l_0 q_{l_0}$ is odd and $ \le k -2$,  
and so the induction hypothesis applies to it.

b) If $l_0$ is odd, $l_0 \ge 5$, and $q_{l_0}$ is even,  
then $k+2 - l_0 q_{l_0}$ is odd and $ \le k-2$,  
we can still apply the induction hypothesis to it.

c)  If $l_0$ is odd, $l_0 \ge 5$, and $q_{l_0}\ge 1$ is odd, then $k+2 - l_0 q_{l_0}$ is even
and $\le k -2$. We  infer that $k+2 - l_0 q_{l_0}  \le  k+2 -5q_{l_0} \le k - 3$. 
We rather use in this situation \eqref{keven}, 
for $\tilde k = k+2 - l_0 q_{l_0} $ is even.

Let us focus on a) and b) and we write
 using our induction hypothesis on $q''$, 
\bea
&&
\Sym[q'] = 
l_0^{q_{l_0}} q_{l_0} ! \, 
\Sym[q''] \leq  l_0^{q_{l_0}} q_{l_0} !   \, 
\Sym[3, 2^{(k+2 - l_0 q_{l_0}  - 3)/2}] \cr\cr
&& 
\le  l_0^{q_{l_0}} q_{l_0} !   \,  3 \cdot 2^{(k+2 - l_0 q_{l_0}  - 3)/2}
((k+2 - l_0 q_{l_0}  - 3)/2)!  \cr\cr
&& 
\le  3 \cdot   2^{(k+2  - 3)/2}  (\frac{ l_0 }{2^{ l_0/2 } })^{ q_{l_0} }
q_{l_0} ! ((k+2 - 3)/2 - q_{l_0}  )!  
\cr\cr
&&
\le 3 \cdot 2^{(k+2  - 3)/2}
((k+2   - 3)/2)!   = \Sym[3, 2^{(k+2  - 3)/2}]
\eea
where  at an intermediate step we use $\frac{l_0 }{ 2 } \ge 2 $, and  $\frac{ l_0 }{2^{ l_0/2 } } \le 1$, 
$l_0 \ge 4$,  the case where $\frac{ l_0 }{2^{ l_0/2 } } = 1$, is precisely
when $l_0 = 4$.

 We deal with the case c).  Note   $l_0 \ge 5$, and $ (l_0 -5 )q_{l_0} \ge 0 $,
 then using \eqref{keven}, we write
 \bea
 &&
 \Sym[q'] = 
l_0^{q_{l_0}} q_{l_0} ! \, 
\Sym[q''] \leq  l_0^{q_{l_0}} q_{l_0} !   \, 
\Sym[2^{(k+2 - l_0 q_{l_0})/2}] \cr\cr
&&
\leq  l_0^{q_{l_0}} q_{l_0} !   \, 
2^{(k+2 -  l_0 q_{l_0} )/2} ((k+2 -( l_0  -2)q_{l_0})/2 -q_{l_0} ) !  \cr\cr
&&
\leq 
2^{(k+2 - 3+ 3  )/2} 
 (\frac{ l_0 }{2^{ l_0/2 } })^{ q_{l_0} }   \, 
  q_{l_0} ! ((k+2 -( l_0  -2)q_{l_0})/2 -q_{l_0} )  !  \cr\cr
  &&
\leq 
3 \cdot 
2^{(k+2 -3 )/2} 
\frac{2^{3/2}}{3}
 (\frac{ l_0 }{2^{ l_0/2 } })^{ q_{l_0} }   \, 
 ((k+2 -( l_0  -2)q_{l_0})/2  )  !  \cr\cr
 &&
 \leq 
3 \cdot 
2^{(k+2 -3 )/2} 
\frac{2^{3/2}}{3}
 (\frac{ l_0 }{2^{ l_0/2 } })^{ q_{l_0} }   \, 
 ((k+2 -3 )/2  )  !  < \Sym[3, 2^{(k+2 -3 )/2}]
 \eea
 that completes the proof of case c). 

  \ 
  
 - Subcase 2:  $q_2>0$ or $q_3>0$. 
 
a) Let us assume that $q_2 >0$, then we write 
$q=[2^{q_2}, q'']$, where $q'' \vdash k+2 - 2q_2 = k+2 -2 \le  k $. 
We  write  $q'' \vdash  k - 2(q_2-1) \le k$
and since $ k - 2(q_2-1)$ is odd,  the induction hypothesis applies to $q''$. 

Then, we obtain 
\bea
&&
\Sym(q) = 2^{q_2} q_2 ! \, \Sym[q'']  \leq 2^{q_2} q_2 ! \,\Sym[3, 2^{(  k - 2(q_2-1) -3)/2} ] \cr\cr
&&
\le  2^{q_2} q_2 ! \, 3 \cdot 2^{(  k - 2(q_2-1) -3)/2} ((  k - 2(q_2-1) -3)/2) ! \cr\cr
&&
\le  3  \cdot 2^{q_2} 2^{(  k + 2 -3)/2-q_2}   q_2 ! \,  ( (  k + 2 -3)/2    -q_2 ) ! \cr\cr
&&
\le \Sym[3, 2^{(  k + 2 -3)/2  )} ] 
\eea 
that is the expression we sought.

b) We now consider the case $q_3>0$. If $q_2 >0$ then we can conclude by the  just above 
argument. Hence, only the situation $(q_2=0 , q_3 >0)$ remains to be dealt with.  Then $q=[3^{q_3}, q'']$
with $q'' \vdash k+2 - 3q_3 \le  k+2 -3 \le  k-1 $ and  the smallest  part
in $q''$ is of minimal size 4. At this moment,  we must study some cases.

b1) If $q_3>0$ is even, then  $q_3\ge 2$ and $q'' \vdash k+2 - 3(q_3-2) -6 =  k-4 - 3(q_3-2)\le k-4$.
Since  $k-4 - 3(q_3-2)$  is odd, the induction hypothesis applies to $q''$. We get
 \bea
 &&
\Sym(q) = 3^{q_3} q_3 ! \, \Sym[q'']  \leq 3^{q_3} q_3 ! \, \,\Sym[3, 2^{(k-4 - 3(q_3-2)-3)/2} ] \cr\cr
&&
\le    3 . 3^{q_3} .2^{(k-1)/2 - 3q_3/2} \,  q_3 !  \, ((k-1 )/2 - q_3 - q_3/2 )! \cr\cr
&&
\le    3 . 3^{q_3} .2^{(k+2-3)/2 - 3q_3/2} \,  q_3 !  \,  \frac{((k+2-3 )/2 - q_3 )! }{ \prod_{i=0}^{q_3/2-1}
((k-1 )/2 - q_3-i)} \crcr
&&
\le    (3 . 2^{- 3/2})^{q_3} \,   \frac{1 }{ \prod_{i=0}^{q_3/2-1}((k-1 )/2 - q_3-(q_3/2-1))} 
\,  \Sym[3,2^{(k+2-3)/2}]\crcr
&&
\le    (3 . 2^{- 3/2})^{q_3} \,   \frac{1 }{ ((k+1 - 3q_3)/2   )^{q_3/2}} 
\,  \Sym[3,2^{(k+2-3)/2}]\cr\cr
&&
\le   (3 . 2^{- 3/2})^{q_3} \, 2^{-q_3/2}  \, \Sym[3,2^{(k+2-3)/2}]
= (3 . 2^{- 2})^{q_3}  \, \Sym[3,2^{(k+2-3)/2}] \cr\cr
&& 
\le   \Sym[3,2^{(k+2-3)/2}]
 \eea
 where at an intermediate step, we use $ k+2 \ge 13$, 
 $k+1 - 3q_3 > 1$. 
 Indeed, since  $k+2 = 3q_3 + |q''|$, where $|q''| \ge 4$ is the sum of parts 
 of $q''$ that is non empty with smallest part at least  $4$. 
 Then 
 \bea
 &&
 0 = k+2  - 3q_3 - (|q''| -4) - 4 =  k-2  - 3q_3 - (|q''| -4) \crcr
 &&
 3  =   k+1  - 3q_3 - (|q''| -4) \le   k+1  - 3q_3 \, , 
 \eea
and since $k+1  - 3q_3 $ is even we have $ k+1  - 3q_3 \ge 4$.

b2) If $q_3>0$ is odd, $q_3 \ge 1$, 
$q'' \vdash k+2 - 3(q_3-1) -3 =  k-1 - 3(q_3-1)\le k-1$. 
There are three subcases to be treated. 
%

 $q_3 = 1$: $q'' \vdash  k-1$, with $k-1$ even so the induction hypothesis  \eqref{keven} applies
to  $q''$, and so we write: 
\bea
\Sym(q) = 3  \, \Sym(q'')  \leq 3 \, \,\Sym[2^{(k-1)/2} ]  = 3 \, \,\Sym[2^{(k+2-3)/2} ]  = \Sym[3, 2^{(k+2-3)/2}]
\eea

$q_3 =  3$: 
$q'' \vdash k+2 - 3*3=  k-7$, with $k-7$ even. Thus, the induction  \eqref{keven} applies
to  $q''$: 
\bea
&&
\Sym(q) = 3^3 3! \;  \, \Sym(q'')  \le 3^3 3! \;  \Sym(2^{(k-7)/2}) 
 = 3^3 3 ! 2^{(k-7)/2}  ((k-7)/2)!  \cr\cr
 &&
 \leq
 3^3 3 ! 2^{(k-1)/2 - 3}  ((k-1)/2 - 3)!    =  
 3^2 3 ! 2^{-3}   \crcr
 && \times \frac{((k-1)/2 - 3)!  ((k-1)/2 - 2)((k-1)/2 - 1)((k-1)/2) }{  ((k-1)/2 - 2) ((k-1)/2 - 1) ((k-1)/2) }  \; (3 \cdot  2^{(k-1)/2 } )  \cr\cr
 &&
    \leq 
 3^2 3 ! 2^{-3}   \crcr
 && \times \frac{1 }{  ((k-1)/2 - 2) ((k-1)/2 - 1) ((k-1)/2) }  \; (3 \cdot  2^{(k-1)/2 } ((k-1)/2)! )   \cr\cr
 &&
  \leq 
 3^2 3 ! 2^{-3}   \crcr
 && \times \frac{1 }{  ((k-1)/2 - 2) ((k-1)/2 - 1) ((k-1)/2) }  \;  \Sym[3, 2^{(k+2-3)/2}]
\eea
Using $k\ge 11$, we have
\bea
\Sym(q)  
 &\leq& 
 3^2 3 ! 2^{-3}  \frac{1 }{  ((11-1)/2 - 2) ((11-1)/2 - 1) ((11-1)/2) }  \;  \Sym[3, 2^{(k+2-3)/2}]\crcr
 & \leq&  
 3^2 3 ! 2^{-3}  \frac{1 }{  (3) (4) (5) }  \Sym[3, 2^{(k+2-3)/2}] < \Sym[3, 2^{(k+2-3)/2}]
\eea
 \ 
 
$q_3 \ge 5$: Then $ k+2 \ge 19$ and 
$q'' \vdash k+2 - 3(q_3-5) -15=  k-13 - 3(q_3-5)\le k-13$. 
Considering that $ k-13 - 3(q_3-5)$ is even, 
we use  the  bound \eqref{keven} on $q''$ and write
\bea
&&
\Sym(q) = 3^{q_3} q_3 ! \, \Sym(q'')  \leq 3^{q_3} q_3 ! \, \,\Sym[2^{(k-13 - 3(q_3-5) )/2} ] 
\cr\cr
&& \le 3 \cdot 2^{(k-13 )/2} (\frac{3^{q_3-1}  }{2^{3(q_3 - 5)/2}})\,   q_3 ! \, ((k-13  )/2 - (q_3-5) - (q_3-5)/2)!
\cr\cr
&&
 \le 3 \cdot 2^{(k-13 )/2} (\frac{3^{q_3-1} }{2^{3(q_3-5)/2}})  \,   q_3 ! \, ((k-1 )/2 -1- q_3- (q_3-5)/2)!
\cr\cr
&&
 \le 3 \cdot 2^{(k-1)/2  -6}(\frac{3^{q_3-1} }{2^{3(q_3-5)/2}}) \,    \    ((k-1  )/2  -1 - (q_3-5)/2 ) )!
 \cr\cr
&&
 \le 3 \cdot 2^{(k-1)/2} \frac{1}{2^6}(\frac{3^{q_3-1} }{2^{3(q_3-5)/2}})  \,  
 \frac{ ((k-1  )/2) !}{ \prod_{i=0}^{  (q_3-5)/2 }  ((k-1  )/2 -  i ) }  
 \cr\cr
&&
\le \frac{1}{2^6}(\frac{3^{q_3-1} }{2^{3(q_3-5)/2}})  \,  
 \frac{ 1}{ \prod_{i=0}^{  (q_3-5)/2   }  ((k-1  )/2 -  i ) }  \; \Sym[3, 2^{(k+2-3)/2}]
 \cr\cr
 &&
\le \frac{1}{2^6}(\frac{3^{q_3-1} }{2^{3(q_3-5)/2}})  \,  
 \frac{ 1}{  ((k-1  )/2 -  (q_3-5)/2  )^{(q_3-5)/2 +1} }  \; \Sym[3, 2^{(k+2-3)/2}]
 \cr\cr
&&
\le  \frac{1}{2^6}(\frac{3^{q_3-1} }{2^{3(q_3-5)/2}})   \,  
 \frac{ 1}{  ((k-q_3  )/2 +2 )^{(q_3-5)/2 +1} }  \; \Sym[3, 2^{(k+2-3)/2}]
\eea
where we used $(k-a)! \prod_{i=0}^{a-1} (k-i)= k!$, $\forall k\ge a\ge 0$. 
We have $k+2 - 3q_3 \ge 4$, therefore $( k - q_3)/2 \ge 4/2 = 2$. 
The above expression finds the bound 
\bea
\Sym(q) 
&\le&    \frac{1}{2^6}(\frac{3^{q_3-1} }{2^{3(q_3-5)/2}})    \frac{1}{ 2^{(q_3-5) +2} } \Sym[3, 2^{(k+2-3)/2}]
 \cr\cr
&\le & (\frac{3^{q_3-1} }{2^{3(q_3-5)/2 + 2q_3/2 +3 }})    \Sym[3, 2^{(k+2-3)/2}]  \cr\cr
&\le&  (\frac{3^{q_3-1} }{2^{(5q_3-9)/2 }})    \Sym[3, 2^{(k+2-3)/2}]
< \Sym[3, 2^{(k+2-3)/2}] 
\eea
 that ends the proof of  the case and of the lemma.

\section{Sage codes for coefficients of the asymptotic series}
\label{app:sage}

The coefficients $a_l$, $l\ge 1$ of the series expansion of $S_{3,m;K}(n)$ as $n \to \infty$
is given by the following program.

We use the  built-in methods

-  \verb|p.centralizer_size()| to compute $\Sym(p)$ for a given partition $p$, 

- \verb|Partitions(k, min_part = p).list() | that produces the list of partitions of $k$, each with parts
larger or equal $q$ involved in the constrained sum $\sum_{ q \vdash  k  : q_1 = q_2 = \cdots = q_p = 0 } $

- and \verb| stirling_number2(l,k) | to evaluate the Stirling number of second kind with 
parameter  $(l,k)$. 

\ 

\noindent{\bf Code for $S_{3,1;K}(n)$ }
\begin{verbatim}
def coeff ( n , l ): 
    som2 = 0 
    m = min(n,l)
    
    for k in range(1,m+1) :    # k = 1 ... m 
        som1 = 0 
        lsk = Partitions(k, min_part = 2).list()    
        for i in range(len(lsk)) : # i = 0 .. len(lsk) -1 
            som1 = som1 + lsk[i].centralizer_size()
        som2 = som2 + som1*stirling_number2(l-1,k-1)
    return som2
\end{verbatim}

\

\noindent{\bf Code for $S_{3,m; K}(n)$}
\begin{verbatim}
def coeffm (m, n , l): 
    som2 = 0 
    
    if m> floor(n/2):
        print "m index out of range"
        return 0
    else: 
        mo = min(floor(n/m),l)
    
    
        for k in range(1,mo+1) :    # k = 1 ... m 
            som1 = 0 
            lsk = Partitions(m*k, min_part = m+1).list()    
            for i in range(len(lsk)) : # i =0 .. len(lsk) -1 
            
                som1 = som1 + lsk[i].centralizer_size()
            som2 = som2 + som1*m^(l-k)*stirling_number2(l-1,k-1)
        
    return som2
    
    \end{verbatim}

\noindent{\bf Code for tabulating the minimal value of $f(n,k)$}

\

In the proof of lemma \ref{fnk}, we use an approximation of $kmin$ the 
minimum of the function $f(n,k)$, for $k\in [K, n]$. Here, we show by numerics 
that, at large and various values of $n$, the following approximation of the minimal 
value of $f(n,k)$ holds: 
\bea
kmin = n - \sqrt{ n } + 1/4
\eea
so that $( n - kmin ) \sim  \sqrt n$ and also $ kmin^{1/2} <  n-k$ holds, for large $n$.

The function \verb|TabMin(nmin , nmax)| 
tabulates the minimal values of $f(n,k)$, for  $n \in [nmin, $ $ nmax]$, 
and $k\in [K=n-13, n]$. It outputs  4-tuples
$$[n , kmin ,  n -  kmin ,   n- \sqrt{n} + 1/4 - kmin ],$$ 
where $kmin$ is the index of the minimum of $f(n,k)$ for $k\in [K=n-13, n]$. 
The claim is that $ n- \sqrt{n} + 1/4 \sim  kmin$,
for large $n$, thenrefore the last entry of the 4-tuple should be small. 
The specific value $14$ is choosen as we run a calculation for 
even $n \in [20,80]$, so that $ \sqrt{n} \in [4,8]$, and therefore 
$kmin \sim n- \sqrt{n} \in  [n-14, n]$. Using this we are investigating
 the neighborhood of $ n- \sqrt{n}$. For odd $n\in [21,79]$, we perform 
 the analogue calculations as well with the hinge value of 13.

\begin{verbatim}

# The function f(n,k) for n and k even 
def feven(n,k):
  return 1.0*factorial (n-k)/factorial(n)* 
        len(Partitions (k , min_part=2).list())*2^(k/2)*factorial (k/2)
 
 
# The function f(n,k) for n and k odd  
def fodd(n,k):
  return 1.0*factorial (n-k)/factorial(n)* 
        len(Partitions (k , min_part=2).list())*2^((k-3)/2)*factorial ((k-3)/2)
 
# Table of values of feven 
def TabEvenCoefksum ( n , K ): 
    cnk =  [ 0 for i in range( ( n - K  )/2 +1 ) ]  
    for i in  range( ( n -  K )/2 +1 ): 
	      ki = K + 2*i 
	      cnk[i] = N(feven( n , ki ))
    return cnk 
    
# Table of values of fodd
def TabOddCoefksum ( n , K ): 
    cnk =  [ 0 for i in range( ( n - K )/2 +1 ) ]  
    for i in  range( ( n -  K  )/2 +1 ): 
    	  ki = K + 2*i 
	       cnk[i] = N(fodd( n , ki))
    return cnk    
    
# Table of values [ n, kmin, n-kmin, n- sqrt (n) + 1/4 - kmin ]
# the parity of nmin will determine which function one chooses 

def TabMin ( nmin , nmax ) : 
    cnk = [ 0 for i in   range ( ( nmax - nmin +2)/2 )  ] 
    for i in  range ( ( nmax - nmin +2)/2 )  : 
    	   n_min = nmin + 2*i
        if nmin%2 == 0 : 
            tab =  TabEvenCoefksum ( n_min  , n_min  - 14 )
            kmin = n_min - 14 +  2*tab.index( min (tab ) )
        if nmin%2 == 1 : 
            tab =  TabOddCoefksum ( n_min  , n_min + 1 - 13 )
            kmin=  ( n_min - 12 ) + 2*tab.index( min (tab ) )
           
        cnk[i] =  [ n_min , kmin,  n_min  -  kmin, 
                    N (( n_min - kmin )  - sqrt (n_min) + 1/4 ) ]  
    
    return cnk 

     
\end{verbatim}
    
- For the case $k$ even we obtain $n\in [20,80]$

\begin{verbatim}
[[20, 16, 4, -0.222135954999580],
 [22, 18, 4, -0.440415759823430],
 [24, 18, 6, 1.35102051443364],
 [26, 20, 6, 1.15098048640722],
 [28, 22, 6, 0.958497377870819],
 [30, 24, 6, 0.772774424948339],
 [32, 26, 6, 0.593145750507619],
 [34, 28, 6, 0.419048105154699],
 [36, 30, 6, 0.250000000000000],
 [38, 32, 6, 0.0855859970310240],
 [40, 34, 6, -0.0745553203367590],
 [42, 36, 6, -0.230740698407860],
 [44, 38, 6, -0.383249580710800],
 [46, 38, 8, 1.46767001687473],
 [48, 40, 8, 1.32179676972449],
 [50, 42, 8, 1.17893218813452],
 [52, 44, 8, 1.03889744907202],
 [54, 46, 8, 0.901530771650466],
 [56, 48, 8, 0.766685226452117],
 [58, 50, 8, 0.634226894136091],
 [60, 52, 8, 0.504033307585166],
 [62, 54, 8, 0.375992125988189],
 [64, 56, 8, 0.250000000000000],
 [66, 58, 8, 0.125961595364039],
 [68, 60, 8, 0.00378874876467883],
 [70, 62, 8, -0.116600265340756],
 [72, 64, 8, -0.235281374238571],
 [74, 66, 8, -0.352325267042627],
 [76, 68, 8, -0.467797887081348],
 [78, 68, 10, 1.41823913367215],
 [80, 70, 10, 1.30572809000084]]
   \end{verbatim}
    
- For the case $k$ odd we obtain $n\in [21,81]$
  
\begin{verbatim}
[[21, 17, 4, -0.332575694955840],
 [23, 19, 4, -0.545831523312719],
 [25, 21, 4, -0.750000000000000],
 [27, 21, 6, 1.05384757729337],
 [29, 23, 6, 0.864835192865496],
 [31, 25, 6, 0.682235637169978],
 [33, 27, 6, 0.505437353461971],
 [35, 29, 6, 0.333920216900384],
 [37, 31, 6, 0.167237469701781],
 [39, 33, 6, 0.00500200160160169],
 [41, 35, 6, -0.153124237432849],
 [43, 37, 6, -0.307438524302000],
 [45, 39, 6, -0.458203932499369],
 [47, 41, 6, -0.605654600401044],
 [49, 43, 6, -0.750000000000000],
 [51, 43, 8, 1.10857157145715],
 [53, 45, 8, 0.969890110719482],
 [55, 47, 8, 0.833801512904337],
 [57, 49, 8, 0.700165564729250],
 [59, 51, 8, 0.568854252131392],
 [61, 53, 8, 0.439750324093346],
 [63, 55, 8, 0.312746066806228],
 [65, 57, 8, 0.187742251701451],
 [67, 59, 8, 0.0646472281275496],
 [69, 61, 8, -0.0566238629180749],
 [71, 63, 8, -0.176149773176359],
 [73, 65, 8, -0.294003745317530],
 [75, 67, 8, -0.410254037844386],
 [77, 69, 8, -0.524964387392123],
 [79, 71, 8, -0.638194417315589],
 [81, 71, 10, 1.25000000000000]]
 \end{verbatim}

\end{document}